\documentclass[letterpaper]{article} %
\usepackage{aaai23}  %
\usepackage{times}  %
\usepackage{helvet}  %
\usepackage{courier}  %
\usepackage[hyphens]{url}  %
\usepackage{graphicx} %
\usepackage{natbib}  %
\usepackage{caption} %
\usepackage{algorithm}
\usepackage{algorithmic}

\usepackage{newfloat}
\usepackage{listings}
\DeclareCaptionStyle{ruled}{labelfont=normalfont,labelsep=colon,strut=off} %
\floatstyle{ruled}
\newfloat{listing}{tb}{lst}{}
\floatname{listing}{Listing}
\usepackage{amsthm}
\usepackage{amsmath}
\usepackage{amssymb}
\usepackage{dsfont}
\DeclareMathOperator*{\argmax}{arg\,max}

\usepackage{graphicx}
\usepackage{todonotes}
\usepackage{subcaption}
\newenvironment{proofsketch}{%
  \proof}{\endproof}

\newtheorem{definition}{Definition}
\newtheorem{theorem}{Theorem}

\usepackage{geometry}
\usepackage{multibib}
\newcites{Appdx}{References}
\setcitestyle{round}

\title{Constrained Submodular Optimization for Vaccine Design}
\author {
    Zheng Dai,\textsuperscript{\rm 1}
    David K. Gifford, \textsuperscript{\rm 1}
}
\affiliations {
    \textsuperscript{\rm 1} Computer Science and Artificial Intelligence Laboratory, Massachusetts Institute of Technology\\
Cambridge, Massachusetts 02139 USA \\
    zhengdai@mit.edu, gifford@mit.edu
}

\begin{document}
\maketitle

\begin{abstract}
Advances in machine learning have enabled the prediction of immune system responses to prophylactic and therapeutic vaccines. However, the engineering task of designing vaccines remains a challenge. In particular, the genetic variability of the human immune system makes it difficult to design peptide vaccines that provide widespread immunity in vaccinated populations.  We introduce a framework for evaluating and designing peptide vaccines that uses probabilistic machine learning models, and demonstrate its ability to produce designs for a SARS-CoV-2 vaccine that outperform previous designs.  We provide a theoretical analysis of the approximability, scalability, and complexity of our framework.
\end{abstract}

\newcommand{\dmr}{diminishing returns}
\newcommand{\Dmr}{Diminishing returns}
\newcommand{\ourmethod}{Optivax-P}
\newcommand{\ourmethodinitials}{OP}
\newcommand{\objf}[1]{\mathcal{F}_{#1}}
\newcommand{\pepset}{\mathcal{P}}
\newcommand{\posrel}{\mathbb{R}^{\geq 0}}
\newcommand{\mhcs}{\mathcal{M}}
\newcommand{\ucon}{\mathcal{F}}
\newcommand{\pcon}{\mathcal{G}}
\newcommand{\bindspmhc}[2]{display(#1,#2)}
\newcommand{\prob}{\text{Pr}}
\newcommand{\SSE}{\textsc{Gap-Small-Set Expansion}}
\newcommand{\califunc}{\mathcal{H}}
\newcommand{\mathbbm}[1]{\mathds{#1}}
\section{Introduction}

Peptide vaccines that expand and activate T cells have emerged as a promising prophylactic and therapeutic approach for addressing health related challenges including infectious diseases and cancer \cite{malonis2019peptide}. In contrast to more conventional live-attenuated vaccines that are based on entire organisms, or subunit vaccines that are based on entire protein subunits, peptide vaccines are based on a small set of protein fragments (peptides) that are sufficient to induce a T cell immune response, enabling the elicitation of far more targeted responses that avoid allergenic and reactogenic responses \cite{li2014peptide}.

The design of a peptide vaccine consists of selecting immunogenic protein fragments, usually referred to as epitopes \cite{li2014peptide}, that when included in a vaccine expand epitope specific T cells. Advances in machine learning have enabled our ability to predict which peptides will be presented by major histocompatibilty complex (MHC) molecules for surveillance by the adaptive immune system \cite{ching2018opportunities, reynisson2020netmhcpan}, which can be used to identify which epitopes will be displayed \cite{sohail2021silico}.  

The epitopes displayed by an individual depend upon the specific alleles of their MHC genes, and thus the peptides displayed by the immune system can vary greatly from individual to individual \cite{zaitouna2020variations}. Therefore, the engineering task of finding a set of peptides that is predicted to be displayed by a large portion of the population remains challenging despite progress on the peptide-MHC display task. 

In this work we introduce a framework for evaluating and designing peptide vaccines that uses probabilistic interpretations of machine learning models, and demonstrate its ability to produce designs for the SARS-CoV-2 vaccine design task that outperform previous designs. We complement this with a theoretical analysis of the approximability, scalability, and complexity of our framework, which may be of independent interest.

\subsection{Our contribution}

To improve the effectiveness of a vaccine it is important to introduce redundancies into its design so the failure of a single displayed peptide to elicit an immune response does not become a single point of failure \cite{ntimescoverage}.  Vaccines designed with an $n$-times coverage objective aim to obtain at least $n$ immunogenic peptide ``hits" in each person.  Having more than one ``hit" provides redundancy to expand multiple T cell clonotypes in an individual to fight disease, protects against peptide sequence drift resulting from pathogen or tumor mutations, protects against the loss of an MHC gene, and accounts for the variability of peptide immunogenicity between individuals. 

We show that optimizing the population coverage for strict $n$-times coverage guarantees cannot be tractably approximated to any constant factor assuming the intractability of the $\SSE$ problem \cite{raghavendra2010graph}, a result which may be of independent interest. We therefore propose a \emph{\dmr{}} framework that uses a soft redundancy guarantee as its objective. The resulting objective is both submodular and monotonic, and can therefore be approximated via a greedy approach which we call \emph{\ourmethod{}}. We supplement the theoretical improvement with an empirical comparison of vaccines designed using our approach and previous designs. Our proposed framework also contributes the following desirable properties: it makes explicit the utility of having redundancy, does not discount the benefits of being covered without redundancy, and is able to reason with uncertainty. We demonstrate how uncertainty values for epitope identification can be derived by calibrating state-of-the-art peptide-MHC display predictors.

While redundancies in a design are important, it is also important that they be \emph{dissimilar redundancies}, since reasons for failure may be shared between similar peptides. This additional constraint that selected peptides be dissimilar is problematic as it allows the problem formulation to encode NP-hard graph problems that in general cannot be approximated to any constant factor. However, by parameterizing on the structure of the constraints, we can derive lower bounds for the performance of the greedy approach which show that the greedy approach can still provide approximation guarantees under certain assumptions.  These bounds may also be of independent interest.

\subsection{Related work}

The use of computational methods to aid vaccine design has taken on an increasingly important role in the vaccine design process over the past two decades \cite{moise2015ivax}. Much of the advancement stems from improvements in the epitope identification task, which has seen impressive improvements with advances in data collection strategies and machine learning \cite{ching2018opportunities,reynisson2020netmhcpan}. While good epitope prediction tools are essential to vaccine design, the focus of this work is on the downstream task of calibrating the predictions and selecting defined epitopes for vaccine inclusion.

Earlier works on vaccine design are reviewed in \citet{oyarzun2015computer}, and employ discrete optimization techniques such as integer linear programming and genetic algorithms to optimize population coverage. However, they do not anticipate or solve the problem of coverage with dissimilar redundancies \cite{ntimescoverage}, which we do in this work. Furthermore, they do not consider the epistemic uncertainty associated with epitope predictions, which we do.

Our work is closely related to the work in \cite{ntimescoverage}, where the use of an objective that accounts for dissimilar redundancies is proposed. However, approximating their proposed objective to any constant factor appears to be an intractable problem, while our objective permits constant factor approximations in polynomial time. Our framework also allows for reasoning about \emph{redundancies} with uncertainty, which theirs does not.

\subsection{Presentation}
In the remaining four sections of the paper we present the optimization problem that we wish to solve (Section 2), provide an algorithm for solving the problem and analyze its runtime and approximation guarantees (Section 3), apply our framework to the SARS-CoV-2 vaccine design problem (Section 4), and conclude with a discussion (Section 5). Theorems are presented where appropriate throughout. Proofs, including intuitive descriptions, are relegated to Appendix \ref{appdx-proofs} for improved flow.
\section{A diminishing returns objective enables theoretical performance guarantees}

Our goal in this section is to formalize the vaccine design problem as an optimization problem. We first show theoretical barriers to obtaining performance guarantees for previous formalizations, and then introduce the \emph{\dmr{}} framework which addresses this.

Peptide vaccines are designed by considering the peptide sequence(s) of a target of interest, for example the proteome of a virus, and selecting a small set of peptides within the target sequences to include in the vaccine. Vaccine peptides are selected such that they elicit an immune response in a large portion of a susceptible population that we wish to vaccinate. This is done by selecting vaccine peptides that are displayed on the cell surface by MHC proteins.  The resulting peptide-MHC complexes activate the cellular immune system. The challenge of selecting a set of peptides arises from the polymorphism present in MHCs within a population. Different MHC alleles have different peptide binding properties, so the peptides must be carefully chosen in order to elicit widespread immune responses from a given population.

\subsection{Preliminaries}

Let $\posrel$ denote non-negative real numbers. Let $E$ be some finite set of elements. Let $F: 2^E \rightarrow \posrel$. We say $F$ is submodular if 
$F(S_1 \cup \{e\})-F(S_1) \geq F(S_2 \cup \{e\})-F(S_2)$ whenever $S_1 \subseteq S_2$ and $e \in E$
, and we say that $F$ is monotonically increasing if $S_1 \subseteq S_2 \implies F(S_1) \leq F(S_2)$ for all $S_1, S_2 \subseteq E$.

Suppose $G = (V,E)$ is a graph. For simplicity, we will at times use $G_V$ to denote its vertex set $V$ and $G_E$ to denote its edge set $E$. The $k$th power of $G$, denoted $G^k$, is defined as the graph $(G^k_V, G^k_E)$, where $G^k_V = G_V$ and $G^k_E$ contains all pairs of vertices between which there exists a path of length less than or equal to $k$ in $G$.

We will use $\mathbbm{1}_X$ to denote an indicator that evaluates to 1 if $X$ is true and 0 otherwise for any proposition $X$.

\subsection{Optimizing population coverage with redundancies is computationally difficult}
\label{sec-ntimescoverdescribe}

It is important for a vaccine to cause the display of multiple epitopes in individuals to provide redundancy in the activation of T cell clonotypes, to expand multiple T cell clonotypes in an individual to fight disease, to protect against peptide sequence drift as a consequence of pathogen or tumor mutations, to protect against the loss of an MHC gene, and to account for the variability of peptide immunogenicity between individuals \cite{ntimescoverage}. In \citet{ntimescoverage}, the authors showed that previous vaccine designs fail to cover significant portions of the population when coverage criteria include these redundancies. To address this, they introduce the $n$-times coverage framework, which involves solving the max $n$-times coverage problem. The problem is defined as follows:

\begin{definition}
Given a ground set, a set of weights over the ground set, a collection of multisets whose elements are from the ground set, and some cardinality constraint $k$, find a collection of $k$ multisets such that the aggregate weights of the elements in the ground set that are covered at least $n$ times is maximized.
\end{definition}

The sum of the weights of the the elements that are covered at least $n$ times is then called the $n$-times coverage. For vaccine design, the ground set corresponds to MHC genotypes, the weights correspond to the percentage of the population with the genotypes, and each multiset corresponds to a peptide, which covers certain genotypes a variable number of times. Solving this problem with cardinality constraint $k$ then gives a vaccine design consisting of $k$ peptides, with the objective that a large portion of the population display at least $n$ peptides (i.e. have at least $n$ peptide-MHC hits).

While this is a natural extension of earlier vaccine design paradigms that do not account for redundancies, it is a computationally difficult problem. The authors have shown in their work that this is an NP-hard optimization problem, and so they propose heuristic approaches. However their proposed approaches have no performance guarantees. Here, we show that this problem is related to $\SSE$, which suggests that finding any constant factor approximation cannot be achieved in polynomial time.

\begin{theorem}
\label{thm-ntimes-approx}
For any $\epsilon > 0$, if there exists a polynomial time algorithm that can achieve an approximation factor of $\epsilon$ to max $n$-times coverage, then there exists a polynomial time algorithm that can decide $\SSE(\eta)$ for some $\eta \in (0,0.5)$.
\end{theorem}

There is currently no known polynomial time algorithm for $\SSE(\eta)$ for any $\eta \in (0,0.5)$. The \emph{Small Set Expansion Hypothesis} conjectures that $\SSE(\eta)$ is NP-hard for any $\eta \in (0,0.5)$, and is currently an open problem related to the Unique Games Conjecture \cite{raghavendra2010graph}.

\subsection{A \dmr{} framework for vaccine design provides a submodular optimization objective}
\label{sec-framework-intro}

A key reason underlying the complexity of the max $n$-times coverage problem is that the utility of a peptide may be hidden until we are close to reaching $n$-times coverage. This makes it difficult select peptides optimally before its utility becomes apparent. To address this, we propose a \dmr{} framework, where peptides will improve the objective at any coverage level. Intuitively, this provides a ``gradient'' along which an optimization procedure can climb.

Formally, let $U:\posrel\rightarrow\posrel$ be some non-negative monotonically increasing concave function such that $U(0) = 0$. Let $\mhcs$ denote the set of MHC genotypes observed in the population. Let $w: \mhcs \rightarrow \posrel$ be a weight function that gives the frequency of each genotype in the population. Let $\pepset$ denote the set of candidate peptides from which a subset is selected for the vaccine design. If $p \in \pepset$ and $m \in \mhcs$, let $\bindspmhc{p}{m}$ denote the predicate of whether $p$ is displayed in an individual with genotype $m$. To model uncertainty, we let $\bindspmhc{p}{m}$ vary over the sample space of some probability space, and we assume that the subset of the probability space where $\bindspmhc{p}{m}$ evaluates to true is always measurable. The objective, parametrized by $U$, can then be written as follows:

\tiny
\begin{equation}
\label{eqn-dr-objective}
    \objf{U}(S) = \sum_{m \in \mhcs} w(m) \; \mathbb{E}[ U( \sum_{p \in S} \mathbbm{1}_{\bindspmhc{p}{m}} ) ]
\end{equation}
\normalsize

Where $S \subseteq \pepset$ is the set of peptides selected for vaccine inclusion. This objective is a monotonically increasing submodular function.

\begin{theorem}
\label{thm-submodular-objective}
For any $U:\posrel\rightarrow\posrel$ that is monotonically increasing and concave, $\objf{U}$ is a monotonically increasing submodular function.
\end{theorem}

As a consequence, we can attain a $(1-e^{-1})$-factor approximation using the greedy approach if no additional constraints are given aside from the cardinality of the peptide set \cite{nemhauser1978analysis}.
Beyond submodularity, this objective contributes the following desirable properties: first, it accounts for the fact that having peptide-MHC hits is useful, even if the redundancy does not obtain a given threshold. While having high redundancy is better than having low redundancy, having low redundancy is better than not displaying any peptides. Second, the utility we expect from attaining a given number of peptide-MHC hits is made explicit through $U$. Third, it allows reasoning with uncertainty by allowing $\bindspmhc{p}{m}$ to be an uncertain event. Many prediction models output a soft classification instead of a hard one, which we can calibrate to attach uncertainties to the classifications.

For an arbitrary distribution over the set of indicator variables $\mathbbm{1}_{\bindspmhc{p}{m}}$ we may need to approximate the expectation in $\objf{U}$ via sampling. However, we can compute the objective $\objf{U}$ exactly and efficiently if we suppose that for a given MHC genotype $m$, the set of indicator variables $\mathbbm{1}_{\bindspmhc{p}{m}}$ are independent. This is almost certainly false given a sufficiently large pool of peptide sequences, since we should be able to significantly improve the performance of a predictor by training it on a sufficiently large number of peptide sequences. However, we can weaken this assumption to $k$-wise independence if we only consider vaccine designs that include at most $k$ peptides. For values of $k$ that are reasonable in the context of designing peptide vaccines, this assumption is more reasonable than the full independence assumption.

Under the independence assumption we can calculate the objective by computing the distribution of the sum via iterated convolutions of Bernoulli distributions, and then taking the expectation using the distribution (see Appendix \ref{appdx-algorithm} for additional details). This runs in time $\mathcal{O}(|\mhcs||S|^2)$, where $|\mhcs|$ is the number of genotypes and $|S|$ is the number of peptides in the vaccine design.

\subsection{Peptide selections need to be constrained to avoid unreasonable designs}
\label{sec-constraints-card-pair}

We impose two types of constraints on the set of peptides selected for vaccine inclusion: a cardinality constraint, and a set of pairwise constraints.

The cardinality constraint is necessary since our objective function is monotonically increasing. Therefore, the full set of candidate peptides $\pepset$ will maximize it. This is undesirable, since peptide vaccines need to be compact to permit effective delivery and to induce effective intolerance in the context of limited immune system capacity.   Therefore, we will impose a cardinality constraint on the set of selected peptides such that it cannot exceed a given size $k$.

The pairwise constraints are required to avoid very similar peptides from being included. Peptide candidates for vaccine inclusion are generated by sliding windows of various sizes across the protein sequence we wish to target. The produces peptides that are highly similar in sequence, such as nested sequences, and including highly related sequences does not truly improve the effectiveness of the vaccine. Furthermore, the assumption that the variables indicating peptide-MHC interactions are independent likely does not hold when peptides are very similar, since it is possible that the predictor makes use of similar features, which result in systematic errors. Therefore, we introduce a set of pairwise constraints $\pcon$ as a graph where the vertex set $\pcon_V = \pepset$, and where edges exist between peptides that are deemed redundant. We then require that the peptides in the vaccine design form an independent set within $\pcon$.

\section{Methods}

\subsection{A greedy approach provides performance guarantees under the \dmr{} framework}

Our goal is the following: given a peptide set $\pepset$, a set of MHC genotypes $\mhcs$, binding credences between all peptides and MHCs, a monotonically increasing concave utility function $U:\posrel \rightarrow \posrel$ with $U(0) = 0$, a cardinality constraint $k$, and pairwise constraints $\pcon$, find a set $S \subseteq \pepset$ that satisfies all the constraints and maximizes the objective function $\objf{U}(S)$. We define the binding credence between a peptide $p$ and an MHC $m$ as the measure of the subset of the probability space where $\bindspmhc{p}{m}$ evaluates to true. Practically, these values behave like probabilities. We use the term credence to emphasize the epistemic nature of the uncertainty.

We present \ourmethod{}, a greedy approach outlined in Algorithm \ref{alg-greedy}, to produce a solution to this problem. The procedure is straightforward: at each iteration we add the peptide that maximally improves the solution to the solution set, then eliminate that peptide and all similar peptides from consideration for all future steps.

\begin{algorithm}[t]
\caption{\ourmethod{}}
\label{alg-greedy}
\begin{algorithmic}
\REQUIRE{A ground set of candidate peptides $\pepset$, a cardinality constraint $k$, a similarity graph $(\pepset,E)$, and a monotone submodular function $F: 2^{\pepset} \rightarrow \mathbb{R}^{\geq 0}$ where $F(\emptyset) = 0$}
\ENSURE{A set $S \subseteq \pepset$ such that $|S| \leq k$}
\STATE{ $S \gets \emptyset$ }
\STATE{ $Q \gets \pepset$ }
\WHILE{$(Q \neq \emptyset) \land (|S| < k)$}
    \STATE{ $x \gets \argmax_{x \in Q} F(S \cup \{x\})$ }
    \STATE{ $S \gets S \cup x$ }
    \STATE{ $N \gets \{ y| \{x,y\} \in E \} \cup \{x\}$ }
    \STATE{ $Q \gets Q\setminus N$ }
\ENDWHILE
\RETURN S\;
\end{algorithmic}
\end{algorithm}

\subsubsection{Runtime analysis of \ourmethod{}}
The naive runtime is $\mathcal{O}(k^3 |\pepset||\mhcs|)$: the objective function is evaluated $|\pepset|$ times to compute the $\argmax$, the $\argmax$ is computed at most $k$ times, and each evaluation of the objective function takes time $\mathcal{O}(k^2|\mhcs|)$ since the designs will never contain more than $k$ elements.

We can improve the runtime by evaluating the marginal improvement rather than the full objective, bringing the overall runtime down to $\mathcal{O}(k^2 |\pepset||\mhcs|)$ (see Appendix \ref{appdx-algorithm}). We can further vectorize the computation to evaluate the $\argmax$ in $\mathcal{O}(1)$ vector operations, reducing the runtime to $\mathcal{O}(k)$ vector operations and $\mathcal{O}(|\pepset|k)$ operations for constraint handling (see Appendix \ref{appdx-algorithm}). However, vector operations require batching if $|\mhcs|$ and $|\pepset|$ are large, so those parameters still play a significant role in the runtime.

Our implementation can generate designs of size $k \approx 10^2$ over a peptide set of size $|\pepset| \approx 10^3$ with $|\mhcs| \approx 10^6$ genotypes in approximately 5 minutes when parallelized over 8 Titan RTX GPUs. See Appendix \ref{appdx-algorithm} for additional details.

\subsubsection{Approximation ratio of \ourmethod{}}
Let $S^*$ denote the true optimum of the optimization problem. If we let each peptide and MHC genotype interact with probability 1 and let $k$ be sufficiently large, then the desired optimization is equivalent to finding the maximum independent set within $\pcon$. Finding any constant factor approximation to max-clique is NP-hard~\cite{zuckerman2006linear}, which then immediately implies that $S^*$ cannot be approximated to any constant factor. Therefore, the quality of the solution produced by \ourmethod{} cannot be unconditionally bounded by a constant factor with respect to $S^*$, since \ourmethod{} runs in polynomial time.

However, we can bound the solution by looking at the graph structure of $\pcon$. Since we are considering cases where similarity relations are mostly generated from sliding windows over linear sequences, we might expect the resulting graph to be of low degree. Let $\Delta(\pcon)$ denote the degree of $\pcon$. Note that in the special case where $\Delta(\pcon) = 0$, there are no pairwise constraints, so the problem reduces to the optimization of a monotonic submodular function under a cardinality constraint. It is well established that the greedy approach attains an approximation ratio of $(1-e^{-1})$ in this case \cite{nemhauser1978analysis}, and that attaining an approximation ratio of $(1-e^{-1}+\epsilon)$ for any $\epsilon > 0$ is NP-hard \cite{feige1998threshold}.

Another property we can look at is the graph power of $\pcon$. We may expect that in the case where a graph looks like a path, taking the graph power would not add too many extra constraints, in which case replacing $\pcon$ with its graph power $\pcon^p$ would not yield a optimum that is too different. Let $S_p^*$ denote the solutions to the more constrained optimizations:

\tiny
\begin{equation}
    S_p^* = \argmax_{\substack{S \subseteq \pepset:\; |S| \leq k\\
    v_1,v_2 \in S \implies \{v_1,v_2\} \notin \pcon^p_E}}
    \objf{U}(S)
\end{equation}
\normalsize

We can then bound the output of \ourmethod{} by incorporating these extra graph parameters:

\begin{theorem}
\label{thm-greedy-ratio}
Let $\hat{S}$ be the output of \ourmethod{}. Then:
\begin{enumerate}
    \item If $\Delta(\pcon) = 0$, then $\objf{U}(\hat{S}) \geq \objf{U}(S^*)(1 - e^{-1})$
    \item If $\Delta(\pcon) > 0$, then $\objf{U}(\hat{S}) \geq \max( \frac{\objf{U}(S_2^*)}{2}, \frac{\objf{U}(S^*)}{1+\Delta(\pcon)})$
\end{enumerate}
\end{theorem}

We can upper bound the best possible performance of polynomial time algorithms.

\begin{theorem}
\label{thm-approx-hard}
Unless P = NP, there exists no polynomial time algorithm that can output an approximation $\hat{S}$ that guarantees either of the following on all inputs for any $\epsilon> 0$:

\begin{enumerate}
    \item $\objf{U}(\hat{S}) \geq \objf{U}(S^*_2) (1 - e^{-1} + \epsilon)$
    \item $\objf{U}(\hat{S}) \geq \objf{U}(S^*)(\frac{1}{1+\Delta(\pcon)})^{(1-\epsilon)} $
\end{enumerate}

\end{theorem}

\subsection{Using machine learning based models to attain binding credences}
\label{sec-credence-formulation}

We define $\bindspmhc{p}{m}$ as a random event in a probability space of beliefs that peptide $p$ is presented by MHC genotype $m$. We use well established state-of-the-art neural network based models (NetMHCpan4.1 and NetMHCpanII4.0 \cite{reynisson2020netmhcpan}) to generate predictions that we use to derive $\prob(\bindspmhc{p}{m})$. We will assume that the derived beliefs are independent (although it is sufficient to assume $k$-wise independence, and only between events that share a genotype - see Section \ref{sec-framework-intro}). While this may not necessarily be true for closely related sequences, we circumvent this by constraining our designs so that they do not contain closely related sequences (see Section \ref{sec-constraints-card-pair}).

There are two classes of MHC molecules that we need to design our vaccine to bind to. However, due to molecular differences between the two molecules the peptides they bind are largely disjoint. We can therefore simplify the problem to that of producing two separate designs, one for Class I MHCs and one for Class II MHCs. NetMHCpan4.1 provides predictions between peptides and Class I MHCs while NetMHCIIpan4.0 provides predictions between peptides and Class II MHCs. For simplicity, we will refer to these predictors together as NetMHCpan.

Having fixed our class of interest, we then define an MHC genotype to be a set of 3-6 distinct alleles. NetMHCpan provides binding predictions between peptides and individual alleles rather than genotypes, so to attain credences for whether a peptide is displayed by a genotype we assume independence between our beliefs that the peptides are displayed by individual alleles and compute the credence as the following:

\tiny
\begin{equation}
    \prob(\bindspmhc{p}{m}) = 1 - \prod_{x \in m} (1-c_{p,x})
\end{equation}
\normalsize

Where $c_{p,x}$ is the credence for binding between a peptide $p$ and a specific allele $x$.

NetMHCpan outputs likelihoods for binding between any given peptide $p$ and any given MHC molecule $x$. However, it is important to ensure that this likelihood is well calibrated. By default, NetMHCpan calibrates itself by comparing the scores it outputs against a repertoire of naturally occurring peptides and classifying a sequence as being displayed if it scores higher than 99.5\% of those peptides when characterizing binding to Class I MHC and 98\% of those peptides when classifying binding to Class II MHC.

To attain well calibrated credences, we make use of publicly available datasets that were used to validate NetMHCpan, which contain no overlap with the dataset used to train NetMHCpan \cite{reynisson2020netmhcpan}. The dataset consists of a set of epitopes, the MHC molecule they bind to, and the natural context they occur in. The epitopes were used as positive samples while the context sequences were used to generate negative samples. The samples were weighted such that the ratio of the weight of all positive samples to that of all negative samples is $1:199$ for Class I MHC and $1:49$ for Class II MHC to match the fraction of natural peptides that NetMHCpan implicitly assumes to be binders.
The samples were then all fed into NetMHCpan to produce predictions. See Appendix \ref{appdx-calibration} for additional details.
The samples were then binned into 20 equally sized bins and the weighted fraction of positive samples within each bin was calculated to produce a calibration curve, which we present in Figure \ref{fig2_calicurve}A.
We then generate a calibration function $\califunc$ by minimizing the following objective:

\begin{figure}[t]
\centering
A)\includegraphics[width=0.82\columnwidth]{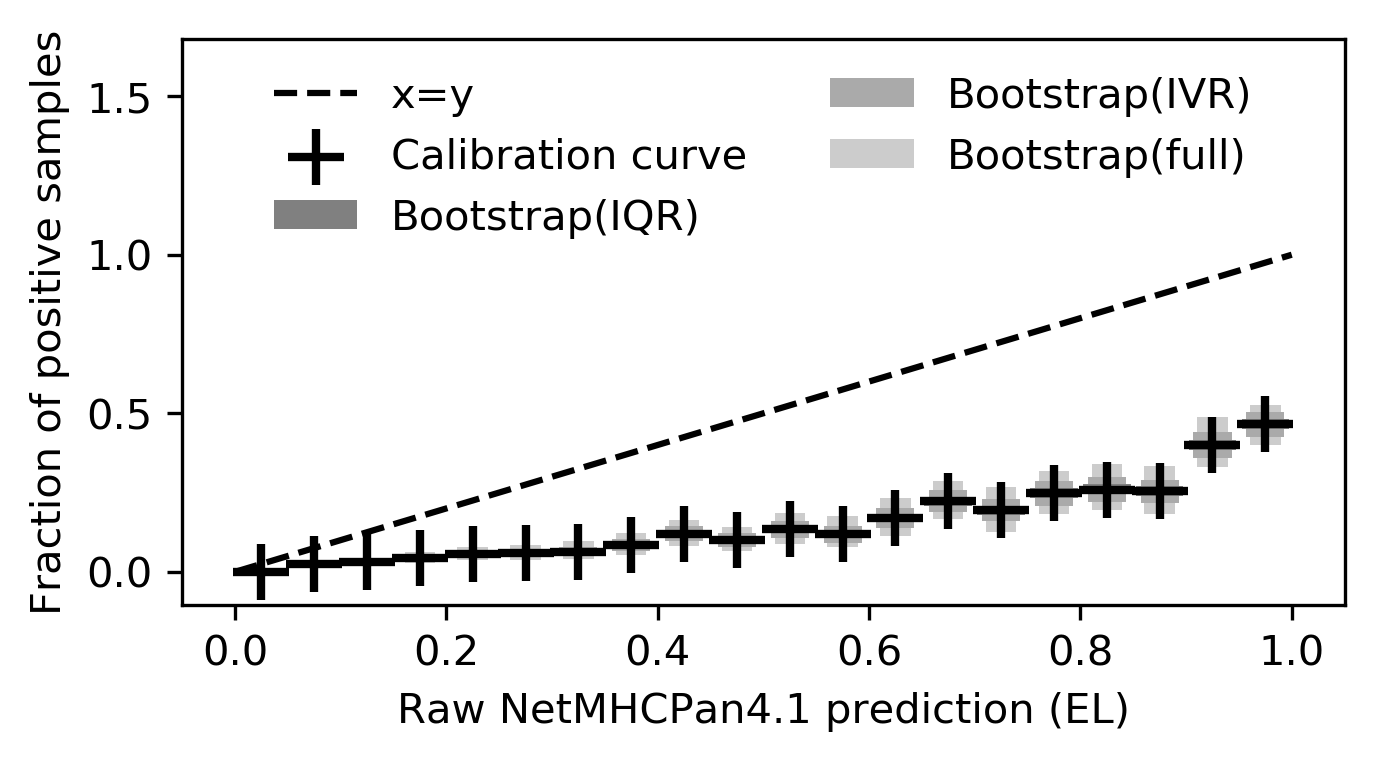}
B)\includegraphics[width=0.82\columnwidth]{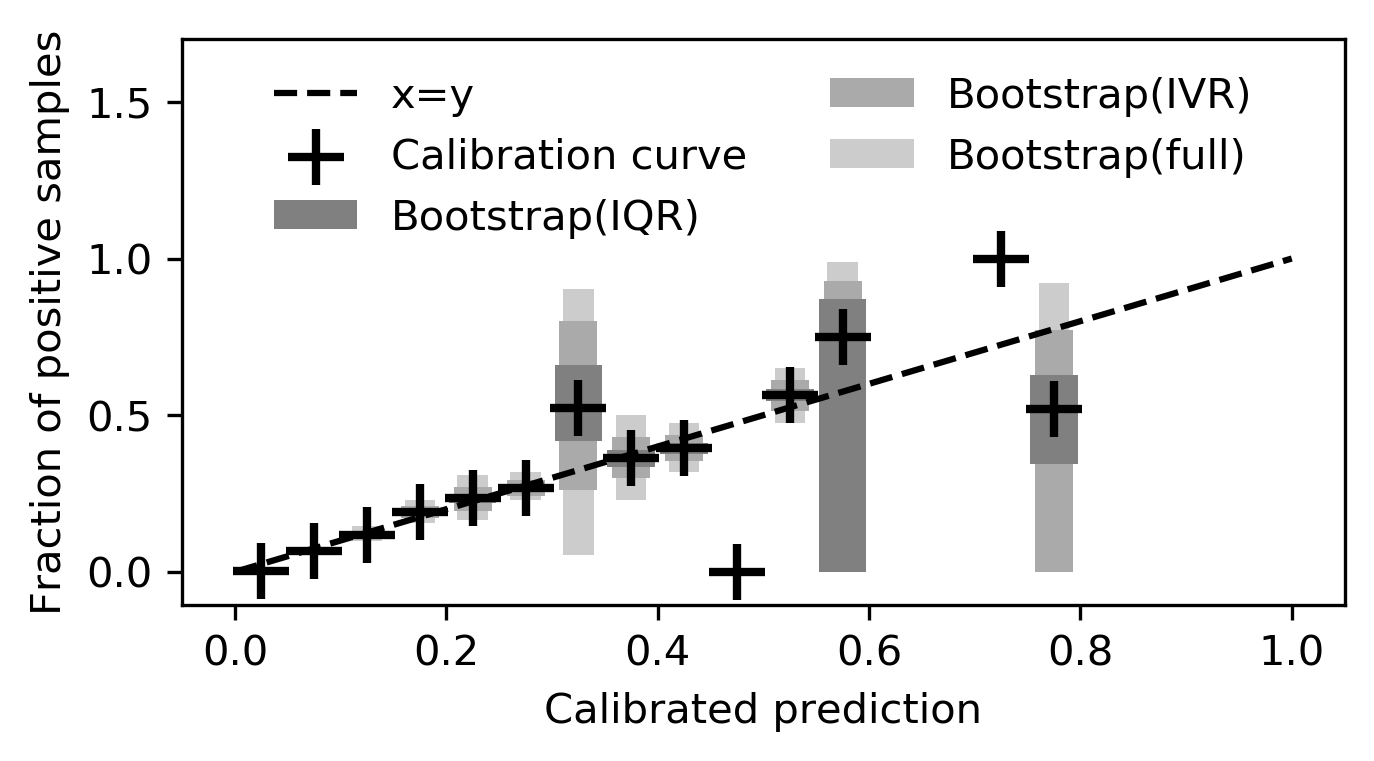}
\caption{Calibration curves for Class I MHC with uncalibrated (A) and calibrated (B) predictions. Matching plots for Class II MHC can be found in Appendix \ref{appdx-calibration}. The curve is made of 20 equally spaced bins between 0 and 1. Populated bins and the fraction of positive samples they contain are indicated by a ``+'', with the surrounding column indicating the interquartile (IQR), interventile (IVR) and full range of a set of 1000 bootstrapped values.}
\label{fig2_calicurve}
\end{figure}

\tiny
\begin{equation}
    \sum_{i=1}^{n-n'} \big( \dfrac{\sum_{j=1}^{n'} w_{i+j}(\califunc(x_{i+j}) - y_{i+j})}{\sum_{j=1}^{n'}w_{i+j}} \big)^2
\end{equation}
\normalsize

Where all samples are indexed by an integer between 1 and $n$, $x_i$ denotes the NetMHCpan predicted value of the sample, $y_i$ is 1 if the sample is positive and 0 if the sample is negative, and $w_i$ is the weight assigned to the sample. The samples are ordered such that $x_i \leq x_{i+1}$ for all $0 \leq i < n$, so $n'$ can be interpreted as the window size. We constrain $\califunc$ to be non-decreasing and non-negative, so this is a form of isotonic regression. Additional details can be found in Appendix \ref{appdx-calibration}.

If NetMHCpan outputs $y$ on an input peptide $p$ and MHC $m$, we set our credence that $p$ binds to $x$ as $\califunc(y)$. A calibration curve of the calibrated predictions of the validation dataset is shown in Figure \ref{fig2_calicurve}B.
\section{Results}
\subsection{Greedy selection outperforms baseline methods}

\ourmethod{} has good worst case theoretical guarantees which are even optimal in the absence of pairwise constraints unless P=NP. Here we empirically check its performance against two baseline approaches in random settings. The first baseline is the \emph{random approach}, where peptides are chosen randomly. The second baseline is via \emph{linear approximation}, where the concave function $U$ in Equation \ref{eqn-dr-objective} is removed to produce a surrogate objective. This surrogate is linear, so it can be trivially optimized to produce a solution.

\begin{figure}[t]
\centering
\includegraphics[width=0.9\columnwidth]{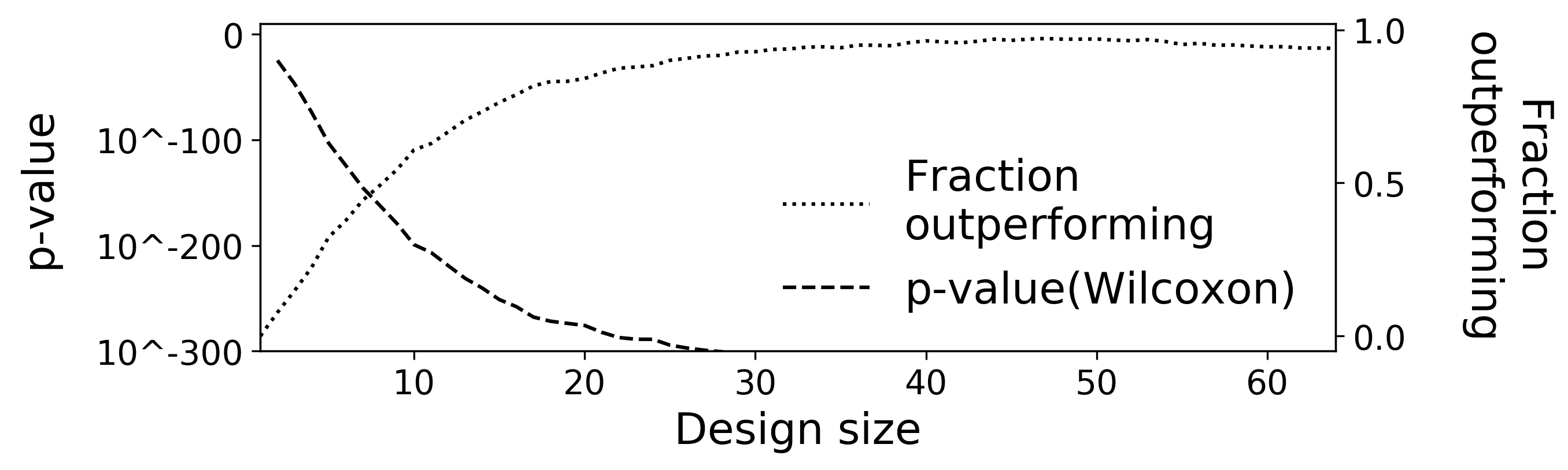}
\caption{For each design size between 1 and 64 inclusive, we compute the fraction of the 2000 settings under which \ourmethod{} outperforms both baselines, and test the hypothesis of whether \ourmethod{} outperforms the best score obtained by the baselines via a one-sided Wilcoxcon signed-rank test. The fraction and p-values (corrected for multiple hypothesis by a factor of 64) are then plotted as a function of the number of peptides selected.}
\label{fig2_synth}
\end{figure}

We compare the algorithms by generating peptide vaccine designs of 64 different sizes in 2000 randomly generated settings, where each setting randomizes the number of genotypes and peptides, the display credences, and the concave function $U$ in the objective $\objf{U}$ (Equation \ref{eqn-dr-objective}). Details can be found in Appendix \ref{appdx-morebench}. We find that \ourmethod{} outperforms the baselines in most settings (Figure \ref{fig2_synth}).

\subsection{Designing a SARS-CoV2 vaccine with \ourmethod{}}

We apply \ourmethod{} in a practical setting by producing vaccine designs for SARS-CoV2. To design our vaccine, we use a set of candidate peptides sourced from \citet{ntimescoverage} which includes peptides from SARS-CoV2 that have been filtered for undesirable properties like high mutation rates, cleavage, and glycosylation. Peptides that were present in the human proteome were also removed, since they may trigger adverse autoimmune responses. \citet{ntimescoverage} have also published a set of genotypes and their frequencies which we use. These frequencies are derived from diverse populations and have been selected to be representative of the global population. We calculated binding credences for each peptide-genotype pair as described in Section \ref{sec-credence-formulation}. Additionally, if a peptide was not present in representative genomes of the Omicron BA.1 and Omicron BA.2 variants of SARS-CoV2 (accession number OM873778 and OW123901 respectively, both retrieved from the The COVID-19 Data Portal \cite{harrison2021covid}) then we set the credence of that peptide being displayed on any MHC molecule to 0.

We applied \ourmethod{} to design vaccines with peptide sets of size 1 and 150 inclusive. Designs were constrained such that no pair of peptides can be within 3 edits (insertions, deletions, or substitutions) of each other for the MHC Class I design, and 5 edits for the MHC Class II design. Designs were optimized for the objective $\objf{U_{T}}$ for $T$ between 1 and 20 inclusive, where $U_T$ is defined as:

\tiny
\begin{equation}
    U_T(x) = \min(x, T)
\end{equation}
\normalsize

\begin{figure}[t]
\centering
A)\includegraphics[width=0.85\columnwidth]{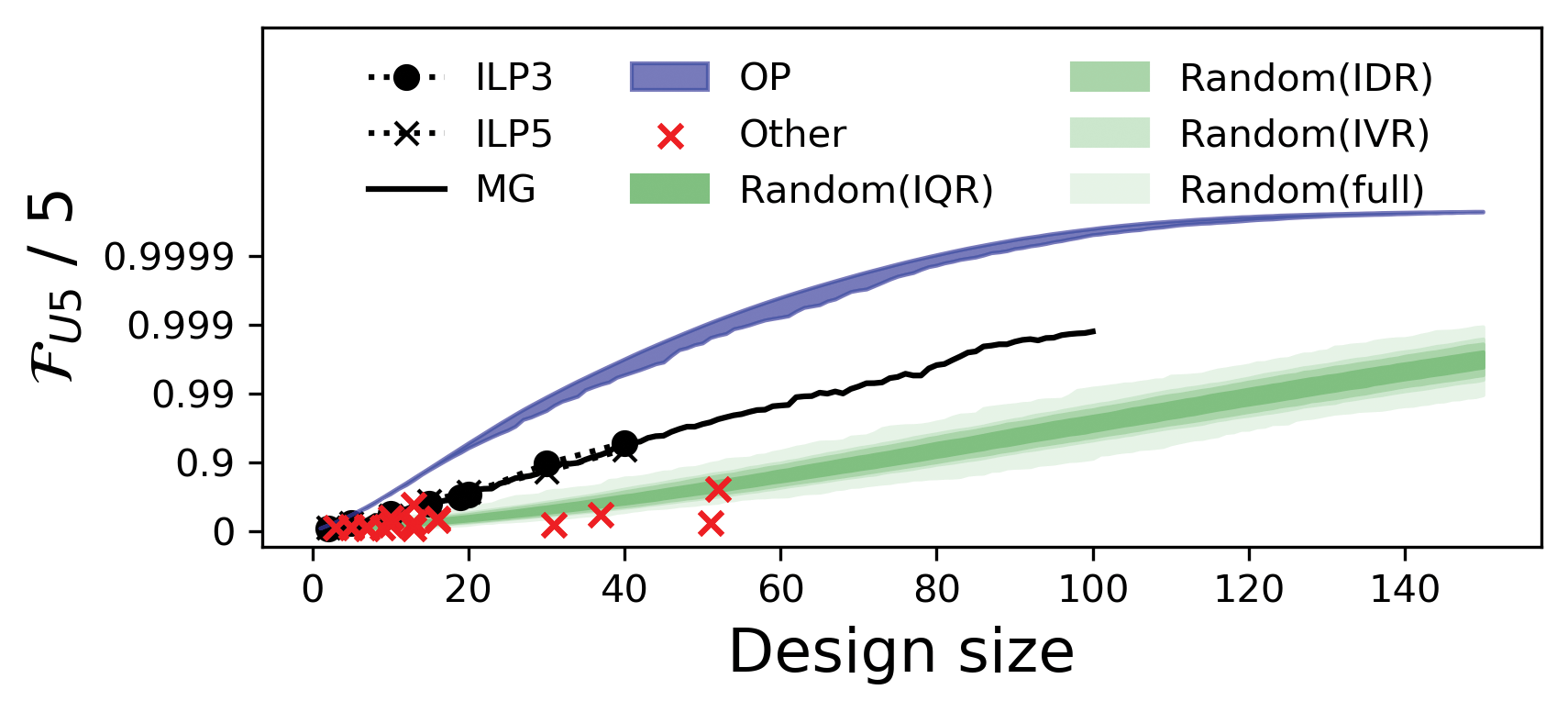}
B)\includegraphics[width=0.85\columnwidth]{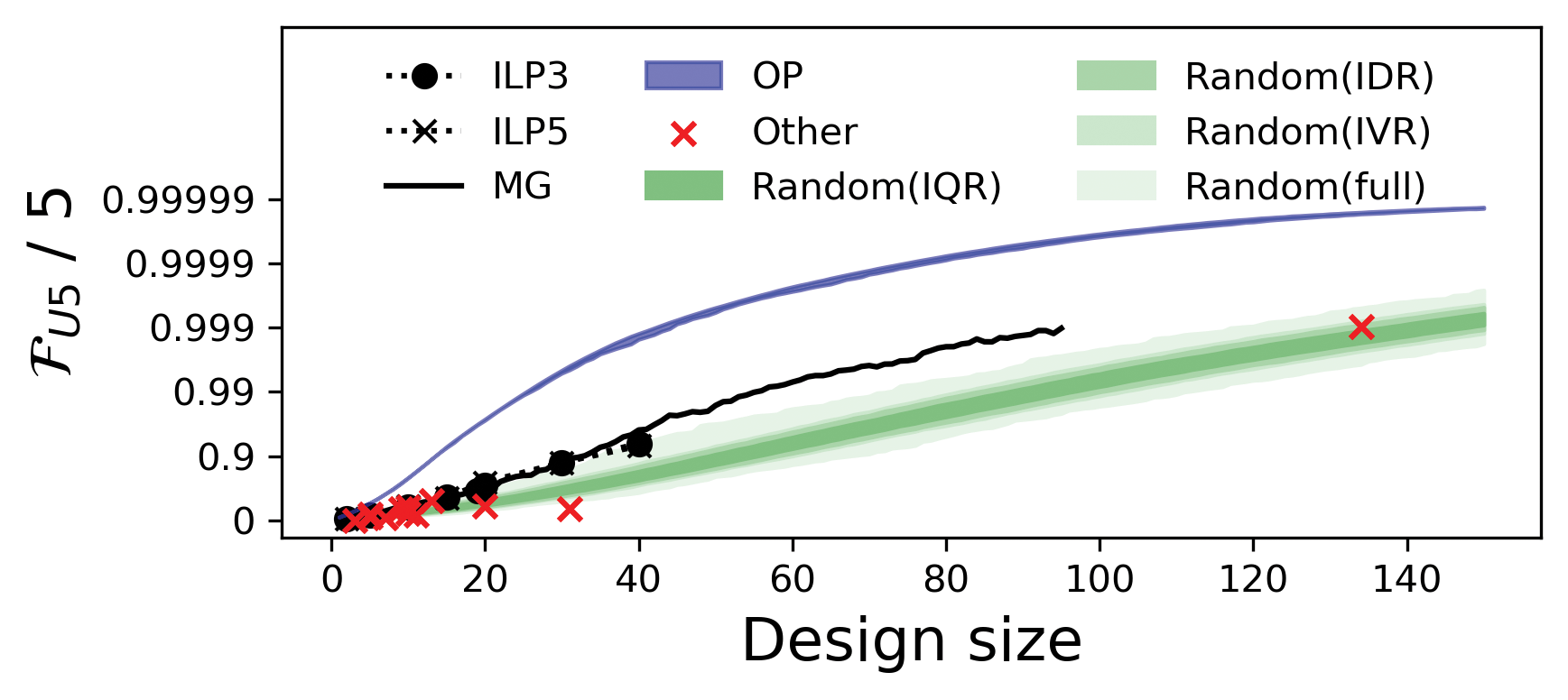}
\caption{For each vaccine design $S$ we compute $\objf{U_{5}}(S)$, divide it by 5, and plot it as a function of $|S|$. Designs for both MHC Class I (A) and MHC Class II (B) are given. For each design size we plot the entire range of designs of that size generated using \ourmethod{} optimized for $\objf{U_{T}}$ with $1 \leq T \leq 20$ (\ourmethodinitials{}), and the interquartile (IQR), interdecile (IDR), interventile (IVR), and full range of 1000 designs of that size sampled uniformly at random. We also plot the locations of the ILP n=3 (ILP3), ILP n=5 (ILP5), and MarginalGreedy (MG) designs from \citet{ntimescoverage} as well as 29 other designs (Other) \cite{abdelmageed2020design,ahmed2020preliminary,akhand2020genome,alam2020design,banerjee2020energetics,baruah2020immunoinformatics,bhattacharya2020development,fast2020potential,gupta2020identification,herst2020effective,lee2020silico,mitramulti,poran2020sequence,ramaiah2021insights,saha2020silico,singh2020designing,srivastava2020structural,tahir2020designing,vashi2020understanding}.}
\label{fig2_comparedesigns}
\end{figure}

$T$ can be viewed as a threshold parameter. This corresponds to a model where each peptide-MHC hit provides incremental protection until a person attains $T$ hits, at which point they are fully protected and stop seeing additional benefits. A comparison between \ourmethod{} designs and previous designs is shown in Figure \ref{fig2_comparedesigns}, where we can see all \ourmethod{} designs score substantially higher on $\objf{U_{5}}$. Evaluations on $\objf{U_T}$ for other $T$ show similar trends and are presented in Appendix \ref{appdx-morebench}.

While this demonstrates that our optimization procedure works and that there are deficiencies in previous designs that are addressed through our designs if our utility model is reflective of reality,
it may not be entirely surprising that vaccine designs optimized for $\objf{U_{T}}$ score well on $\objf{U_{T}}$, even for mismatching $T$. To control for this, we also evaluate designs produced by \ourmethod{} on the $n$-times coverage objective proposed by \citet{ntimescoverage} and described in Section \ref{sec-ntimescoverdescribe}. We modify our credences such that they are only 0 or 1, and such that they closely match the values used by \citet{ntimescoverage} (i.e. $\prob(\bindspmhc{p}{m})=1$ if and only if the genotype $m$ is present within the peptide $p$, where peptides are viewed as multisets of genotypes in the max $n$-times coverage framework).

\begin{figure}[t]
\centering
A)\includegraphics[width=0.95\columnwidth]{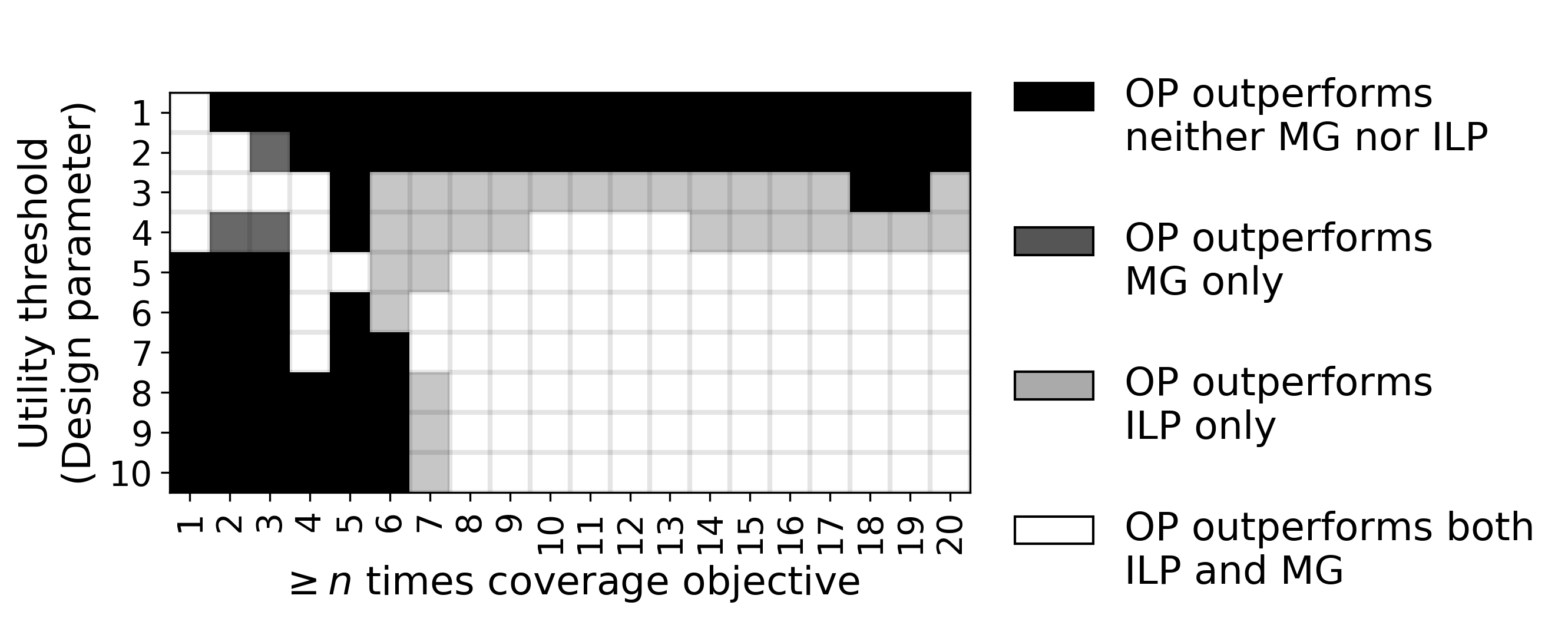}
B)\includegraphics[width=0.95\columnwidth]{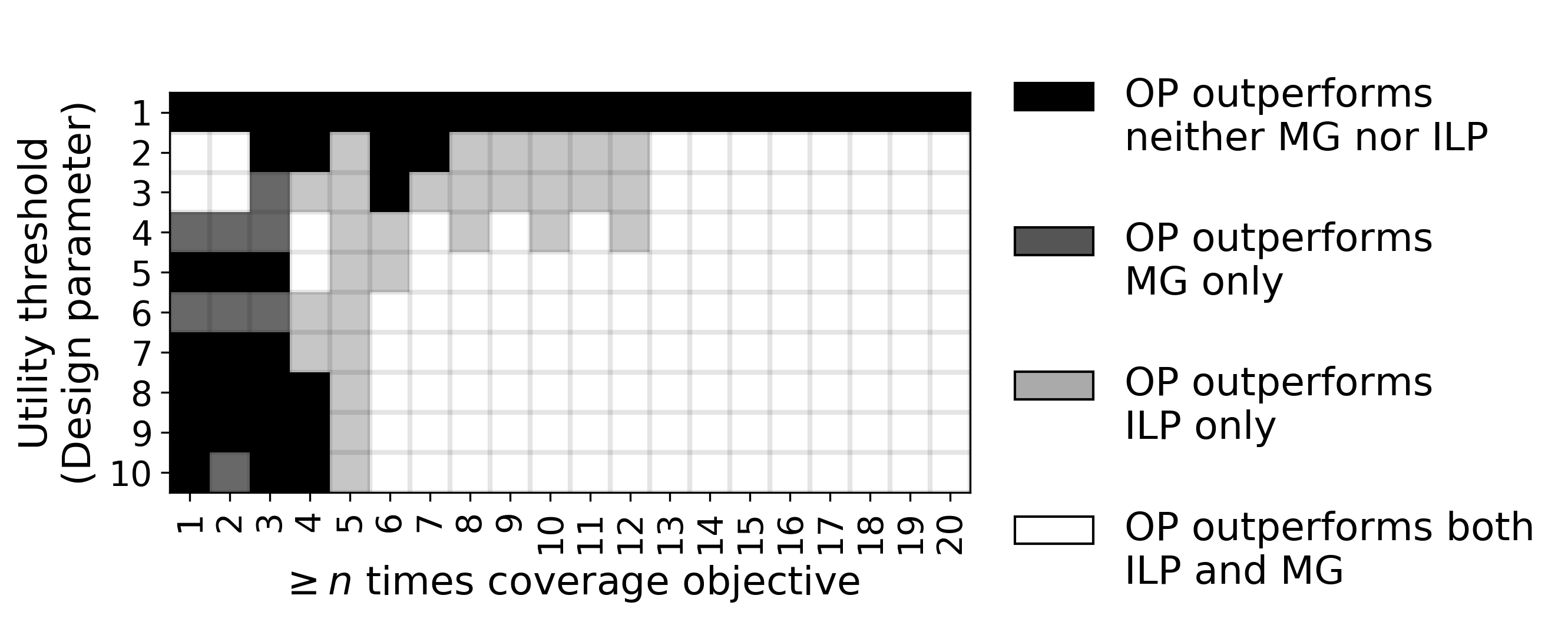}
\caption{We compare designs generated by \ourmethod{} (\ourmethodinitials{}) to designs generated by the ILP and MarginalGreedy (MG) approaches from \citet{ntimescoverage} using the $n$-times coverage objective. A white square at position $(x,y)$ indicates that the design proposed by \ourmethod{} optimized against $\objf{U_y}$ outperforms both the ILP and MarginalGreedy design when evaluated with $x$-times coverage. A black square indicates that the \ourmethod{} design was outperformed by the ILP or MG design, and a gray square indicates that the \ourmethod{} design was outperformed by either the ILP or MG design. Comparisons for both MHC Class I (A) and MHC Class II (B) are given.}
\label{fig2_compareevalvax}
\end{figure}

We then produced designs of size 19 using $U_T$ for all $T$ between 1 and 10 inclusive. These designs were then assigned a score equal to their $n$-times coverage for $n$ between 1 and 20 inclusive. In Figure \ref{fig2_compareevalvax} we compare these scores to the scores attained by designs generated by \citet{ntimescoverage}, which also each contain 19 peptides. We see that \ourmethod{} is highly competitive even when evaluated on a objective separate from the one it was optimized for.

\section{Discussion}

We have introduced the diminishing returns framework and \ourmethod{} for designing epitope based peptide vaccines. The framework is based on constrained submodular optimization, which permits \ourmethod{} to provide performance guarantees despite the NP-hardness of the problem, unlike previous approaches. We also show how we can probabilistically interpret the outputs of machine learning models to allow reasoning with uncertainty within the framework. Finally, we demonstrated that \ourmethod{} achieves superior performance against past vaccine designs on the SARS-CoV-2 vaccine design task, and achieves comparable performance even when evaluated against previous objectives.

\ourmethod{} is highly scalable, allowing us to optimize over potentially millions of candidate peptides for a single vaccine. The ability to reason with uncertainty also gives us a much richer language for expressing the properties of a peptide: for instance, instead of filtering out peptides prone to mutation, we can consider the probability of sequence drift. These factors allow us to consider a far wider range of peptide vaccine design tasks, which we plan to use in the future.

\section*{Acknowledgements}
This work was supported in part by Schmidt Futures and a C3.ai grant. The authors would also like to thank Ge Liu, Alexander Dimitrakakis, Brandon Carter, and members of the Gifford lab for useful discussions and feedback.

\bibliography{references}
\onecolumn
\appendix

\section{Proofs}
\label{appdx-proofs}
\subsection{Proof of Theorem \ref{thm-ntimes-approx}}

\begin{proofsketch}
We can encode a graph into the $n$-times coverage problem. To do this, we let each peptide represent a vertex and we let each genotype represent an edge. A genotype displays a peptide (i.e. a peptide contains a genotype under the multiset formulation) if and only if the edge it represents is incident to the peptide. If we require 2-times coverage, then an edge is ``covered'' only if both vertices at its endpoints are included in the selected set.

Therefore, selecting a set of peptides that cover a lot of genotypes is equivalent to finding a set of vertices that have a lot of edges between them. If the graph is regular, this is equivalent to finding a non-expanding set. This then allows us to reduce from $\SSE$.
\end{proofsketch}

For reference, we state the definition of $\SSE(\eta)$ adapted from \citeAppdx{raghavendra2010graph}:

\begin{definition}
Given a $d$-regular graph $G = (V,E)$ and some integer $k$, determine which of the following is the case:

\begin{enumerate}
    \item There exists $S \subseteq V$ with $|S| = k$ such that the number of edges leaving $S$ is less than $dk\eta$.
    \item For all $S \subseteq V$ where $|S| = k$, the number of edges leaving $S$ is more than $dk(1-\eta)$.
\end{enumerate}
\end{definition}

This is a promise problem: if neither is true then we are free to output whatever we want.

Suppose that there exists a polynomial time algorithm that approximates $n$-times coverage to an approximation ratio of $\epsilon > 0$ (i.e. if $OPT$ is the value of the true solution, the solution returned by the algorithm attains value $\epsilon OPT$).

Then there exists an $\eta \in (0,0.5)$ such that we can solve $\SSE(\eta)$: given an input $d$-regular graph, let each vertex represent a peptide, and let each edge represent a genotype. A peptide is displayed by a genotype if and only if the vertex represented by that peptide is incident to the edge represented by that genotype. We then compute an approximation for max 2-times coverage using our algorithm with cardinality constraint $k$.

Let $\hat{S}$ be the value of the solution returned by the algorithm, and let $S^*$ be the value of the optimal solution. Let $\hat{E}$ be the number of genotypes covered by $\hat{S}$ and let $E^*$ be the number of genotypes covered by $S^*$.

Let $f(x) = dk - 2x$. Since each vertex has degree $k$, and $\hat{E}$ and $E^*$ are the number of edges that have both endpoints in $\hat{S}$ and $S^*$ respectively, $f(\hat{E})$ and $f(E^*)$ are the number of edges that leave $\hat{S}$ and $S^*$ respectively.

Let $p$ be some sufficiently large value that is greater than 0 such that $\epsilon(1-\epsilon^p) > \epsilon^p$ and such that $\epsilon^p < 0.5$.
Output case 1 if $f(\hat{E}) < dk(1-\epsilon^p)$, and output case 2 otherwise.

This procedure completes in polynomial time, and we claim that this correctly solves $\SSE(\epsilon^p)$. To verify this, suppose that we are in case 1. We then have:

\begin{align}
    f(E^*) &< \epsilon^p\\
    dk(1-\epsilon^p) &< 2E^*
\end{align}

Since $\hat{S}$ attains an approximation ratio of $\epsilon$, we have:

\begin{align}
    f(\hat{E}) &\leq f(\epsilon E^*)\\
    &< f(\epsilon(dk(1-\epsilon^p))\\
    &= dk(1 - \epsilon(1-\epsilon^p))\\
    &< dk(1-\epsilon^p)
\end{align}

The last line follows since we defined $p$ to be large enough for that to hold. Since $f(\hat{E}) < dk(1-\epsilon^p)$, the algorithm indeed outputs case 1.

Suppose instead that we are in case 2. But then $f(\hat{E}) > dk(1-\epsilon^p)$ because all sets of size $k$ have more than $dk(1-\epsilon^p)$ edges leaving it. Therefore, the algorithm correctly outputs case 2.

Therefore, this procedure decides $\SSE(\epsilon^p)$ in polynomial time.

\subsection{Proof of Theorem \ref{thm-submodular-objective}}

\begin{proofsketch}
Since $U$ is concave, the marginal utility of each peptide diminishes as number of bound peptides increase when looking at a specific genotype, given a particular realization of which peptides are displayed and which ones do not. The objective function from Equation \ref{eqn-dr-objective} can be viewed as a weighted average of the marginal utilities that has been averaged over each genotype and each possible realization of which peptides bind and which ones do not. Since the marginal utilities are individually diminishing, a convex combination of those utilities must also be diminishing.
\end{proofsketch}

Let $S \subseteq T \subsetneq \pepset$, and let $e \in U \setminus T$. Let $S' = S \cup \{e\}$ and let $T' = T \cup \{e\}$. It suffices to show that $\objf{U}(S') - \objf{U}(S) \geq \objf{U}(T') - \objf{U}(T)$. We have the following:

\begin{align}
    \objf{U}(S') - \objf{U}(S) &= \sum_{m \in \mhcs} w(m) \; \mathbb{E}[ U( \sum_{p \in S'} \mathbbm{1}_{\bindspmhc{p}{m}} ) - U( \sum_{p \in S} \mathbbm{1}_{\bindspmhc{p}{m}} ) ]\\
    &= \sum_{m \in \mhcs} w(m)
    \sum_{u \in \mathcal{U}}\prob(u)
    \big(
    U(H(S',u)) - U(H(S, u))
    \big)
\end{align}

Where $\mathcal{U}$ denotes the set of all possible outcomes (i.e. all possible truth assignments to the predicates $\bindspmhc{p}{m}$ for all $p \in \pepset$ and $m \in \mhcs$), and $H(S,u)$ denotes the number of hits found in set $S$ under outcome $u$. $\prob(u)$ denotes the probability assigned to outcome $u$.

Since $S \subseteq T$, it must be the case that $H(S,u) \leq H(T,u)$ for any outcome $u \in \mathcal{U}$. Since $e$ is either a hit or not a hit under outcome $u$, $H(S',u) - H(S,u) = H(T',u) - H(T,u) = H(\{e\}, u)$ for any outcome $u \in \mathcal{U}$.

When these two conditions are paired with the observation that $U$ is concave, we have:

\begin{align}
    \objf{U}(S') - \objf{U}(S) &=  \sum_{m \in \mhcs} w(m)
    \sum_{u \in \mathcal{U}}\prob(u)
    \big(
    U(H(S',u)) - U(H(S, u))
    \big)\\
    &\geq
    \sum_{m \in \mhcs} w(m)
    \sum_{u \in \mathcal{U}}\prob(u)
    \big(
    U(H(T',u)) - U(H(T, u))
    \big)\\
    &= \objf{U}(T') - \objf{U}(T)
\end{align}

Which then implies that $\objf{U}$ is submodular.

\subsection{Proof of Theorem \ref{thm-greedy-ratio}}

\begin{proofsketch}
Suppose we have our greedy solution, and suppose we then ask an oracle for what the true solution is. If we take the union of these two solutions then by monotonicity of the objective this combined set scores at least as well as the true optimum.

We now note that the reason the greedy procedure did not pick the members of the true solution was because they either add less marginal gain than any of the elements that were picked, or because they were removed from consideration at some point by the choice of some element in the greedy solution. We then note that each element can only block a limited number of other elements, and if the greedy algorithm chose that element over the elements it blocked then it must have had greater marginal utility. This then allows us bound the amount by which the true solution can augment the greedy solution, which in turn bounds the approximation ratio of the greedy solution.
\end{proofsketch}

If $\Delta(\pcon) = 0$, then there are not constrains, and it is well established that the greedy approach indeed attains an approximation ratio of $(1-e^{-1})$\citeAppdx{nemhauser1978analysis}.

Suppose otherwise. Let $e_1, e_2, ..., e_a$ be the elements chosen by Algorithm \ref{alg-greedy} in that order. Let $o_1, o_2, ..., o_b$ be the elements of $S^*_2 \setminus \hat{S}$. 
Pair an $e_i$ an $o_j$ if $e_i$ shares an edge with $o_j$. If a single $o_j$ shares edges with multiple $e_i$, choose the pair randomly. A single $e_i$ cannot share edges with more than one $o_i$, since otherwise that would imply that there exists a path of length two between those two $o_i$ that passes through the $e_i$.

Suppose some $o_j$ are left over. We then pair them up randomly with the remaining $e_i$. If there are unpaired elements, it must be the $e_i$, since otherwise the unpaired $o_j$ would have been added to the greedy solution.

Now consider the trajectory of adding elements to a solution in the order $e_1, e_2, ..., e_a, o_1, o_2, ..., o_b$. By monotonicity, this must attain a value at least as great as $S^*_2$.

Suppose $e_i$ is paired with $o_j$. Then the marginal improvement of taking $e_i$ must have been better than the marginal improvement of taking $o_j$, since otherwise $o_j$ would have been chosen instead of $e_i$. Therefore, the score attained by $e_1, e_2, ..., e_a, o_1, o_2, ..., o_b$ cannot be more than 2 times larger than the score attained by $e_1, e_2, ..., e_a$. But since the value attained by $S^*_2$ can be no greater than $e_1, e_2, ..., e_a, o_1, o_2, ..., o_b$, it must be the case that the value attained by $S^*_2$ is no more than 2 times larger that the value attained by the greedy solution. This establishes the following:

\begin{equation}
    \objf{U}(\hat{S}) \geq \frac{\objf{U}(S_2^*)}{2}
\end{equation}

Similarly, let $q_1, q_2, ... q_b$ be the elements of $S^* \setminus \hat{S}$. We similarly associate $e^i$ with $q_j$ if they share an edge, and if a $q_j$ shares edges with multiple $e_i$ the association is chosen arbitrarily. Note that no $e_i$ can by associated with more than $\Delta(\pcon)$ $q_j$.

We associate the remaining $q_j$ with $p_i$ such that no $p_i$ is associated with more than $\Delta(\pcon)$ $q_j$. If there are leftovers they must belong to $\hat{S}$ because otherwise they would have been included in the greedy solution.

Now consider the trajectory of adding elements to a solution in the order $e_1, e_2, ..., e_a, q_1, q_2, ..., q_b$. By monotonicity, this must attain a value at least as great as $S^*$.

Suppose $e_i$ is associated with $q_j$. Then the marginal improvement of taking $e_i$ must have been better than the marginal improvement of taking $q_j$, since otherwise $q_j$ would have been chosen instead of $e_i$. Since each $e_i$ can be associated with up to $\Delta(\pcon)$ $q_i$, the score attained by $e_1, e_2, ..., e_a, q_1, q_2, ..., q_b$ can be no larger than $\Delta(\pcon)+1$ times the value attained by $\hat{S}$. But then that means $S^*$ does not attain a value that is more than $\Delta(\pcon)+1$ times the value of the greedy solution. This establishes the following:

\begin{equation}
    \objf{U}(\hat{S}) \geq \frac{\objf{U}(S^*)}{1+\Delta(\pcon)}
\end{equation}

Combining the two inequalities then yields the desired statement:

\begin{equation}
    \objf{U}(\hat{S}) \geq \max( \frac{\objf{U}(S_2^*)}{2}, \frac{\objf{U}(S^*)}{1+\Delta(\pcon)})
\end{equation}

\subsection{Proof of Theorem \ref{thm-approx-hard}}

\begin{proofsketch}
The first guarantee cannot be met because \citetAppdx{feige1998threshold} showed that a special case of our optimization (max $k$-cover) cannot be approximated to within the given factor.

The first guarantee cannot be met because \citetAppdx{zuckerman2006linear} showed that a special case of our optimization (max clique) cannot be approximated to within the given factor.
\end{proofsketch}

We restate the two approximation guarantees that cannot be given if $P \neq NP$:

\begin{enumerate}
    \item $\objf{U}(\hat{S}) \geq \objf{U}(S^*_2) (1 - \frac{1}{e} + \epsilon)$
    \item $\objf{U}(\hat{S}) \geq (\frac{\objf{U}(S^*)}{1+\Delta(\pcon)})^{(1-\epsilon)} $
\end{enumerate}

To establish the first inapproximability result, we can encode an instance of max $k$-cover in our optimization problem: let each genotype represent a member of the ground set and let each peptide represent an element of the set system. We let the genotype display a peptide with probability 1 if the element represented by the peptide is contained in the set represented by the genotype, and with probability 0 otherwise. Define the concave utility function as $U(x) = \min(x,1)$, and let $\pcon$ be a set of unconnected vertices, so $\objf{U}(S^*) = \objf{U}(S^*_2)$. If each genotype is given weight 1, then $\objf{U}(S)$ gives the number of elements covered by $S$. The first inapproximability result then follows from the inapproximability of max $k$-cover to a ratio of $(1 - \frac{1}{e} + \epsilon)$ for any $\epsilon > 0$ \citeAppdx{feige1998threshold}.

To establish the second inapproximability result, we can encode an instance of independent set in our optimization problem: let each peptide represent a vertex, and for each peptide introduce a genotype that displays that peptide with probability 1 and displays all other peptides with probability 0. Define the concave utility function as $U(x) = \min(x,1)$. If each genotype is given weight 1, then $\objf{U}(S)$ gives the number of elements contained in $S$. If we set $\pcon$ to be the instance of indepenent set we wish to encode, then $\objf{U}(S^*)$ is the size of the largest independent set. The second inapproximability result then follows from the fact that independent set cannot be approximated to a ratio of $|\pcon_V|^{1-\epsilon}$ for any $\epsilon > 0$ \citeAppdx{zuckerman2006linear}, and from the fact that $1+\Delta(\pcon) \leq |\pcon_V|$.

\newpage
\section{Calibration}
\label{appdx-calibration}

\subsection{Data processing}

Epitope data that were used to evaluate the performance of NetMHCpan4.1 and NetMHCIIpan4.0 was acquired from servers linked to by \citetAppdx{reynisson2020netmhcpan}. The dataset consists of a set of triplets containing the epitope sequence, context sequence, and the HLA that the epitope binds to. The dataset has been filtered so that it contains no overlap with the training set of NetMHCpan4.1 and NetMHCIIpan4.0 \citeAppdx{reynisson2020netmhcpan}.

We first completely remove the epitope from the context sequence, potentially chopping the context into multiple chunks. We then slide a window with the same length as that of the epitope into the context chunks, producing multiple peptides that match the length of the epitope. The epitope is then assigned a weight of 1, while the remaining peptides are assigned weights such that the sum of those weights is 199 for Class I and 49 for Class II. This is to reflect the rule-of-thumb that within a natural distribution of peptides, the top scoring 0.5\% (for Class I) and 2\% (for Class II) of peptides are strong binders, which is the default rule used by NetMHCpan4.1 and NetMHCIIpan4.0 to classify peptides.

Each sequence was then fed to NetMHCpan4.1 or NetMHCIIpan4.0 for to produce binding likelihoods to their associated HLA. The epitope is given a ground truth value of 1 and the peptides derived from the contexts are given a ground truth value of 0. This potentially underestimates the number of binders, since it is possible that peptides in the context sequences also bind. However, our credences maintain soundness: it should not be the case that a vaccine design that is unlikely to work will appear to be effective with high probability under these credences.

Our processing transforms the Class I MHC dataset of 1660 (epitope, HLA, context) triplets into 1645988 (prediction, ground truth, weight) triplets, and the Class II MHC dataset of 917 (epitope, HLA, context) triplets into 1201559 (prediction, ground truth, weight) triplets.

\subsection{Calculating calibration curves}

Given a set of (prediction, ground truth, weight) triplets, we compute the calibration curve by binning the triplets by their prediction values. We use 20 bins spaced equally between 0 and 1. For each bin that is populated, we calculate the weighted average of its ground truth values:

\begin{equation}
    \dfrac{ \sum_i w_i*y_i }{ \sum_i w_i }
\end{equation}

\begin{figure}
\centering
A)\includegraphics[width=0.45\columnwidth]{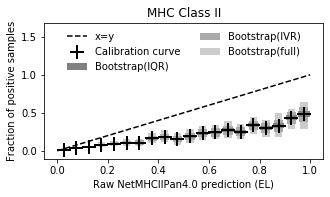}
B)\includegraphics[width=0.45\columnwidth]{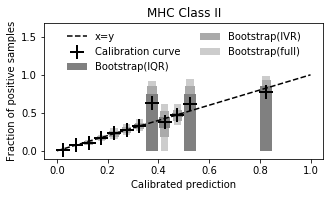}
\caption{Calibration curves for Class II MHC with uncalibrated (A) and calibrated (B) predictions. Populated bins and the fraction of positive samples they contain are indicated by a ``+'', with the surrounding column indicating the interquartile (IQR), interventile (IVR) and full range of a set of 1000 bootstrapped values.}
\label{fig2_appdx_calicurve}
\end{figure}

Where $i$ indexes the peptides within a bin and $w_i$ and $y_i$ denote the weight and ground truth value of peptide $i$ respectively. The weighted average of a bin is a statistic computed from a set of triplets, so we quantify the uncertainty of that statistic using 1000 bootstrapped samples of the triplets of that bin. Figure \ref{fig2_calicurve} gives calibration curves for Class I MHC, while matching curves for Class II MHC are provided in Figure \ref{fig2_appdx_calicurve}.

\subsection{Generating the calibration function}

We use a variant of isotone regression to generate a calibration function $\califunc$ as described in Section \ref{sec-credence-formulation}. Given a list of $n$ triplets $(x_i, y_i, w_i)$, where $x_i$ is raw predicted value of peptide $i$, $y_i$ is the ground truth value of peptide $i$, and $w_i$ is the weight assigned to peptide $i$, ordered such that for all $1 \leq i < n$ we have $x_i \leq x_{i+1}$, we generate $\califunc$ by minimizing the following function, subject to the constraint that $a \leq b \implies \califunc(a) \leq \califunc(b)$ and $\forall a \; \califunc(a) \geq 0$.

\begin{equation}
\label{eqn-appdx-isotonicRegression}
    \sum_{i=1}^{n-n'} \big( \dfrac{\sum_{j=1}^{n'} w_{i+j}(\califunc(x_{i+j}) - y_{i+j})}{\sum_{j=1}^{n'}w_{i+j}} \big)^2
\end{equation}

Implementation-wise, what we do is to slide a window of size $n'$ across a list of triplets sorted by prediction outputs to produce a pair of weighted averages of calibrated predictions and weighted averages of ground truth values. For our use case we take $n'$ to be 1000. We then calculate the squared error and perform projected gradient descent with respect to the squared error over the calibrated weighted predictions. Projection is used to maintain the monotonicity of the calibrated predictions.

This optimization defines $\califunc$ over the finite domain of raw predicted values. Let $X$ be the set of all raw predicted values. We use the following criterion to extend these predictions to $\mathbb{R}$.

\begin{equation}
    \califunc(x) = \max\bigg( \{ \califunc(x) | x' \leq x, x' \in X \} \cup \{0\} \bigg)
\end{equation}

While the objective is strictly convex with respect to the averaged values, there are potentially multiple sets of values that produce the same set of averaged values, leading to potentially multiple solutions that optimize Equation \ref{eqn-appdx-isotonicRegression}. We therefore fix the first $n'-1$ calibrated predictions to be 0, which is reasonable since it turns out that the first $999$ ground truth values are also 0. Having fixed these values, there is now a bijective linear function between the remaining $n-n+1'$ calibrated prediction values and the set of $n-n+1'$ averaged calibration values. Therefore, the objective is also strictly convex with respect to the unaveraged calibrated prediction values.

\newpage
\section{Implementation details}
\label{appdx-algorithm}

\subsection{Computing $\objf{U}$}

To compute $\objf{U}(S)$ in time $\mathcal{O}(|\mhcs| |S|^2)$, it suffices to calculate $\mathbb{E}[ U( \sum_{p \in S} \mathbbm{1}_{\bindspmhc{p}{m}} )$ in time $\mathcal{O}(|S|^2)$ for each $m \in \mhcs$ since we can then add up all the values.

To calculate $\mathbb{E}[ U( \sum_{p \in S} \mathbbm{1}_{\bindspmhc{p}{m}} )$ in time $\mathcal{O}(|S|^2)$, we use iterated convolutions. We store a zero indexed list $D$ of size $|S|+1$ initialized such that $D_i=0$ except for $i=0$, where we have $D_0 = 1$. Then for each $p \in S$, we loop in reverse order from $i=|S|$ to $i=0$ and update the list by setting $D_i$ to $ (1-x)D_{i} + (x)D_{i-1}$, where $x$ is the credence that $p$ is displayed by $m$. $D_0$ is set to $(x)D_0$. The result is a convolution of a distribution with probability mass function $D$ a Bernoulli random variable whose chance of success is $x$. In other words, after each iteration of the loop the value of $D_i$ represents the probability of attaining $i$ hits from a vaccine given the peptides that have already been looped over. Performing this loop for all members of $S$ then gives the probability distribution for $\sum_{p \in S} \mathbbm{1}_{\bindspmhc{p}{m}}$ if all random variables in the sum are independent. We can then take the expectation by using the identity $\mathbb{E}[f(Y)] = \sum_y f(y)\prob(Y=y)$. Each loop runs in time $\mathcal{O}(|S|)$, $\mathcal{O}(|S|)$ loops need to be run, so computing the distribution takes time $\mathcal{O}(|S|^2)$ overall. If the utility function is provided in some data structure that allows random access (e.g. an array), then the expectation can be computed in time $\mathcal{O}(|S|)$. Thus,  $\mathbb{E}[ U( \sum_{p \in S} \mathbbm{1}_{\bindspmhc{p}{m}} )$ can be calculated in time $\mathcal{O}(|S|^2)$, which gives an overall runtime of $\mathcal{O}(|\mhcs| |S|^2)$ for calculating $\objf{U}(S)$.

\subsection{Computing the marginal improvement}

Let $S$ be fixed, and suppose $S' = S \cup \{e\}$ for some $e \in \pepset$. Then we can calculate $\objf{U}(S') - \objf{U}(S)$ in time $\mathcal{O}(|\mhcs| |S|)$ instead if we store the distribution of $\sum_{p \in S} \mathbbm{1}_{\bindspmhc{p}{m}}$ for each $m \in \mhcs$. This is because we only need to convolve a single Bernoulli random variable over the distribution of $\sum_{p \in S} \mathbbm{1}_{\bindspmhc{p}{m}}$ to get the distribution of $\sum_{p \in S'} \mathbbm{1}_{\bindspmhc{p}{m}}$. This then allows us to calculate the expectation in time $\mathcal{O}(|S|)$, which then gives an overall runtime of $\mathcal{O}(|\mhcs| |S|)$ for calculating the marginal improvement.

At the end of each greedy step, we can update the distributions of $\sum_{p \in S} \mathbbm{1}_{\bindspmhc{p}{m}}$ to $\sum_{p \in S'} \mathbbm{1}_{\bindspmhc{p}{m}}$ for all $m \in \mhcs$ in time $\mathcal{O}(|\mhcs| |S|)$ by performing a single convolution between the distribution and a Bernoulli random variable.

\subsection{Vectorization}

The calculation of the marginal differences can be vectorized: let $D'$ be the probability mass function of $\sum_{p \in S'} \mathbbm{1}_{\bindspmhc{p}{m}}$ and let $D$ be the probability mass function of $\sum_{p \in S} \mathbbm{1}_{\bindspmhc{p}{m}}$. Let $x$ be the credence that the element in $S' \setminus S$ is displayed on $m$. Let $D$ and $D'$ be represented by vectors where $D_i$ is the probability that the random variable distributed with probability mass function $D$ takes on value $i$, and likewise for $D'$. Let $U$ be a vector representing the utility function, where $U_i = U(i)$. Then the expected value of $U(Y)$ is $U\cdot D$ if $Y$ is distributed with probability mass function $D$ (likewise for $D'$).

If $Z$ is a vector, let $Sh(Z)$ denote a shift operation where $Sh(Z)_i = Z_{i-1}$ and $Sh(Z)_0 = 0$. Let $Sh^{-1}(Z)$ be the reverse operation where $Sh(Z)_i = Z_{i+1}$ and $Sh(Z)_n = 0$ if $n$ is the size of $Z$. If we ensure that the entry at the last index of $D$ is 0 such that $Z \cdot D = Sh(Z) \cdot Sh(D)$ and $Sh^{-1}(Sh(D)) = D$, then the marginal difference is:

\begin{align}
    \mathbb{E}[U(\sum_{p \in S'} \mathbbm{1}_{\bindspmhc{p}{m}})] - \mathbb{E}[U(\sum_{p \in S} \mathbbm{1}_{\bindspmhc{p}{m}})] &= U\cdot D' - U\cdot D
\end{align}

This can be rearranged as the following:
    
\begin{align}
    U\cdot D' - Sh(U) \cdot Sh(D)
    &= U \cdot ( (x)Sh(D) + (1-x)D ) - Sh(U) \cdot Sh(D)\\
    &= U \cdot (x)Sh(D) - (x) Sh(U) \cdot Sh(D)\\
    &= (x)(U - Sh(U)) \cdot ( sh(D) )\\
    &= (x)(Sh^{-1}(U) - U) \cdot D
\end{align}

Therefore, if we precompute and vectorize $(Sh^{-1}(U) - U)$, the marginal difference is simply a scaled dot product. This is a vector operation. We can then further tensor over $\mhcs$ and $\pepset$ to parallelize most of the operations.

\subsection{Details on hardware, software, and runtime}

Our experiments were all run on Titan RTX GPUs, each with 24190MiB of memory. Our algorithms were all implemented in Python 3.7.3 and make use of PyTorch 1.7.1 and Numpy 1.18.5, and were all run on Ubuntu 18.04.3 LTS.

\begin{figure}
  \begin{center}
        \begin{subfigure}[c]{.95\textwidth}
        \includegraphics[width=\textwidth]{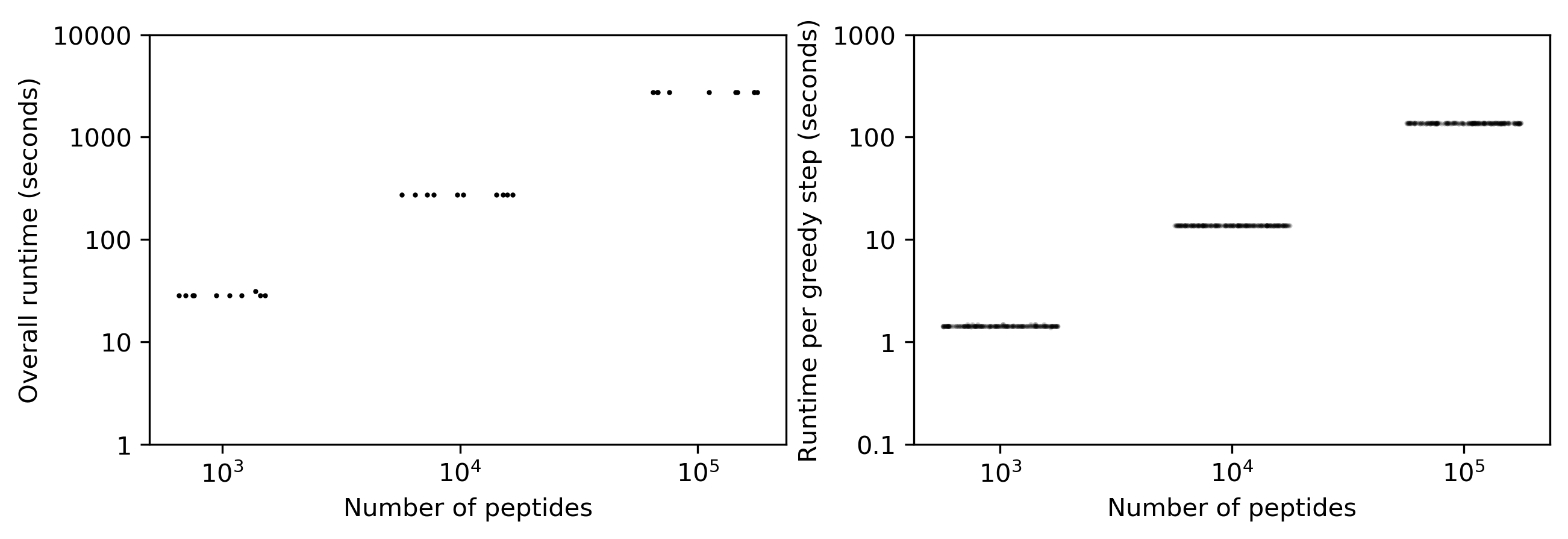}
        \caption{Overall runtimes of \ourmethod{} on one Titan RTX GPU are provided in strip plots on the left panel. Each box is generated from 10 executions. Overall runtimes for each greedy iteration on one Titan RTX GPU are provided on the right panel, where is box is generated from 200 runtimes (20 iterations for 10 runs). The y-axis is presented on a log scale. Some instability in the runtime arises when running on small numbers of peptides because startup and preprocessing times are significant at those timescales.}
        \end{subfigure}
        \begin{subfigure}[c]{.95\textwidth}
        \includegraphics[width=\textwidth]{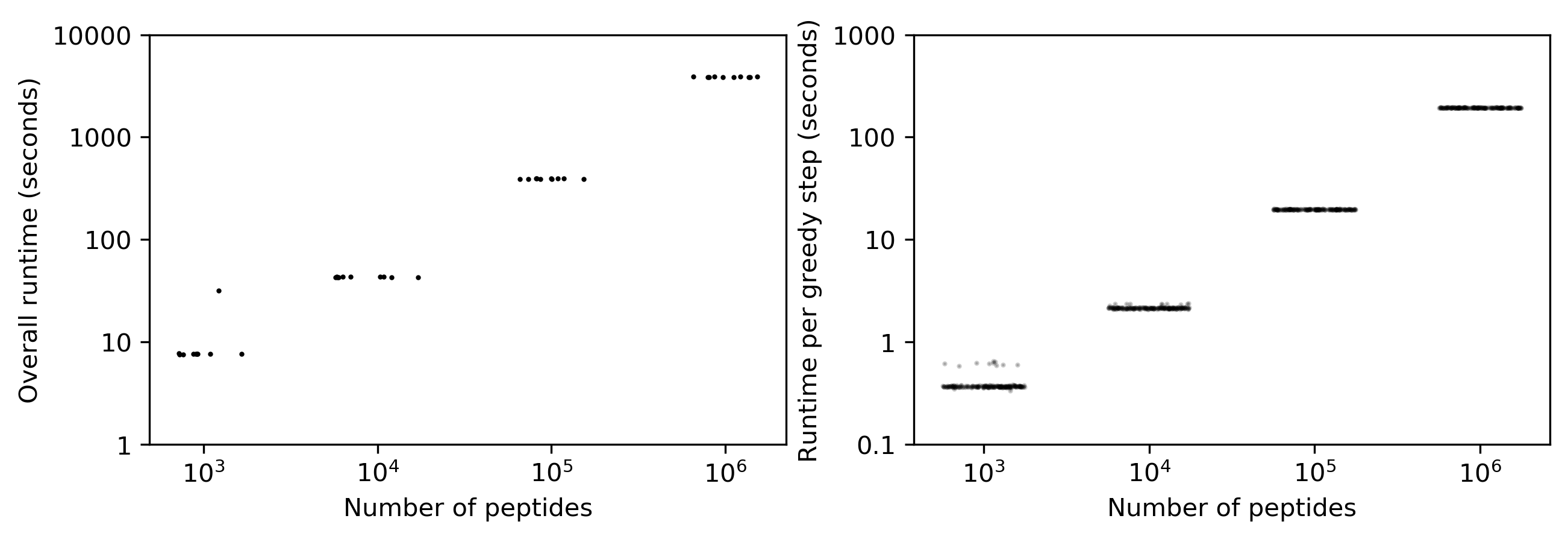}
        \caption{Overall runtimes of \ourmethod{} on 8 Titan RTX GPUs are provided in strip plots on the left panel. Each box is generated from 10 executions. Overall runtimes for each greedy iteration on 8 Titan RTX GPUs are provided on the right panel, where is box is generated from 200 runtimes (20 iterations for 10 runs). The y-axis is presented on a log scale.}
        \end{subfigure}
        \begin{subfigure}[c]{.95\textwidth}
        \includegraphics[width=\textwidth]{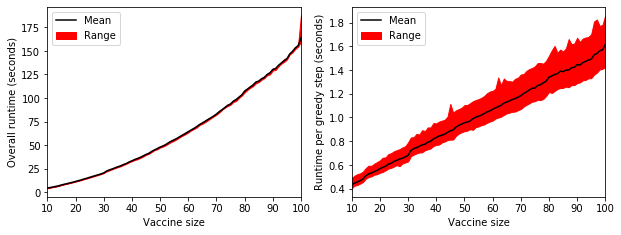}
        \caption{Average runtimes over 10 executions of \ourmethod{} on 8 Titan RTX GPUs are plotted in the left panel, parameterized by the size of the vaccine design. The range of runtimes are also given. Average runtimes for each greedy iteration on 8 Titan RTX GPUs are provided on the right panel, along with the observed ranges.}
        \end{subfigure}
  \end{center}
  \caption{Runtime benchmarks}
  \label{fig-supp-runtime}
\end{figure}

We benchmarked the runtime of Algorithm \ref{alg-greedy} by generating vaccine designs of size 20 using synthetic datasets where binding credences are drawn uniformly and independently between 0 and 1, with $|\mhcs| = 10^6$ and varying sizes of $|\pepset|$. 10 runs were excecuted for each $|\pepset|$. We ran the benchmarks both parallelized over 8 GPUs. We also ran the benchmarks using a single GPU to check how much parallelization over multiple GPUs helps.
The runtimes we recorded are given in Figure \ref{fig-supp-runtime}A and \ref{fig-supp-runtime}B.

Furthermore, we fixed $|\pepset|$ to $10^3$, and generated vaccine designs of size between 10 and 100 inclusive. The runtimes are given in Figure \ref{fig-supp-runtime}C.

\newpage
\section{Benchmarking details}
\label{appdx-morebench}

\subsection{Generating random settings for optimization}

We generate 2000 settings in which to benchmark our algorithms against baselines. The number of peptides, the number of genotypes, the genotype weights, the credences assigned to whether a given genotype displays a given peptide, and the optimization objective $\objf{U}$ were all randomized independently and were drawn in the following way:

\begin{enumerate}
    \item The number of peptides $|\pepset|$ and number of genotypes $|\mhcs|$ were drawn uniformly from integers between 512 and 2048 inclusive.
    \item All genotype weights were drawn independently and identically from the exponential distribution and then normalized so that they sum to 1. Note that the scale parameter of the distribution is irrelevant since the values are normalized.
    \item A value $\alpha$ is set to be $0.005*100^X$, where $X$ is drawn from the uniform distribution supported between 0 and 1.
    \item For each $p \in \pepset$ and $m \in \mhcs$, $\Pr(\bindspmhc{p}{m})$ is assigned $X_{p,m}^{\alpha}$, where the values for $X_{p,m}$ are drawn independently and identically from the uniform distribution supported between 0 and 1.
    \item The function $U$ parametrizing $\objf{U}$ was set to be the following for all integer values:
    \begin{equation}
        U(x) = \sum_{i=1}^{x} \sum_{j=i}^{10} Z_j
    \end{equation}
    Where for $1 \leq j \leq 10$, the $Z_j$ are drawn independently and uniformly from the lognormal distribution with location parameter 0 and scale paramter 2. For simplicity we define $\sum_{i=a}^b t = 0$ if $a > b$ regardless of $t$. This ensures that $U$ has decreasing second difference and a non-negative first difference, which in turn ensures that $U$ is monotonically increasing and concave (i.e. has monotonically decreasing first differences).
\end{enumerate}

For each setting, designs of size 1-64 inclusive were generated using Algorithm \ref{alg-greedy}. Baseline designs of size 1-64 inclusive were chosen by greedily selecting peptides that were displayed on the most (weighted) genotypes without regards to potentially diminishing returns (linear approximation). A second set of baseline designs were chosen by sampling uniformly at random without replacement to produce a trajectory of 64 designs of size 1-64 inclusive (random). We show in the main body of this work that Algorithm 1 outperforms both approaches (see Figure \ref{fig2_synth}).

\begin{figure}
\centering
A)\includegraphics[width=0.45\textwidth]{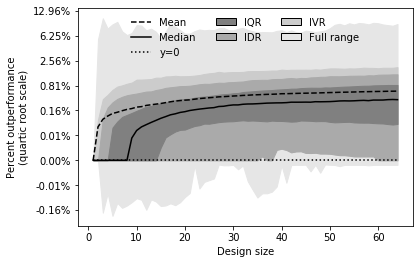}
B)\includegraphics[width=0.45\textwidth]{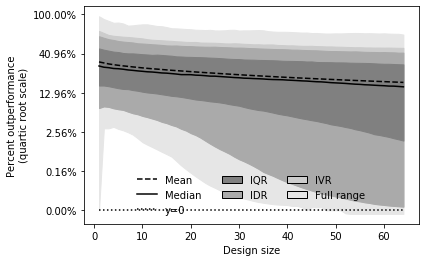}
\caption{We compare baseline designs against designs generated by Algorithm \ref{alg-greedy} on 2000 randomly generated settings. We compute the amount by which Algorithm \ref{alg-greedy} outperforms a baseline via \emph{percent outperformance}, where we subtract the score attained by Algorithm \ref{alg-greedy} by the score attained by the baseline, and then divide by the score attained by Algorithm \ref{alg-greedy}. We then plot for each design size the interquartile (IQR), interdecile (IDR), interventile (IVR), and full range of the distribution of percent outperformances, as well as the mean and median of the distribution. The scores are plotted on a quartic root scale (i.e. the y position of a point with a y value of $y'$ is $|y'|^{1/4} * sgn(y')$, where $sgn$ denotes the sign function). We compare the outputs of Algorithm \ref{alg-greedy} against the linear approximation baseline in (A), and against the random baseline in (B). }
\label{fig2_appdx_synth}
\end{figure}

We also calculated the amount by which Algorithm \ref{alg-greedy} outperforms the baseline, which we present in Figure \ref{fig2_appdx_synth}. The linear approximation baseline is much more competitive than the random baseline, although Algorithm \ref{alg-greedy} still outperforms it by more than $10\%$ in some settings.

\subsection{Analysis of \dmr{} designs on additional utility scores}

\begin{figure}
\centering
A)\includegraphics[width=0.45\textwidth]{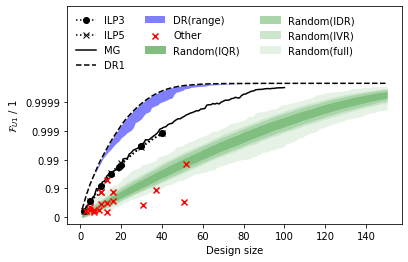}
E)\includegraphics[width=0.45\textwidth]{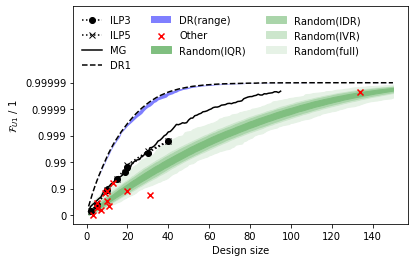}
B)\includegraphics[width=0.45\textwidth]{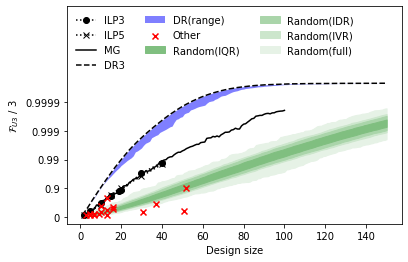}
F)\includegraphics[width=0.45\textwidth]{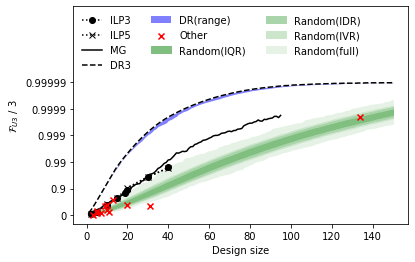}
C)\includegraphics[width=0.45\textwidth]{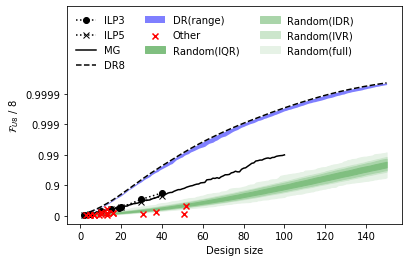}
G)\includegraphics[width=0.45\textwidth]{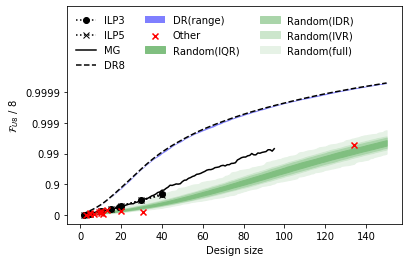}
D)\includegraphics[width=0.45\textwidth]{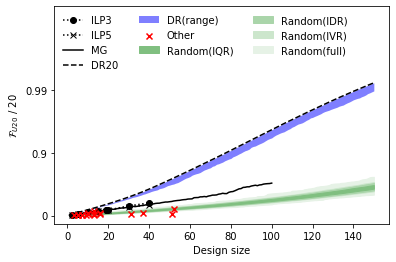}
H)\includegraphics[width=0.45\textwidth]{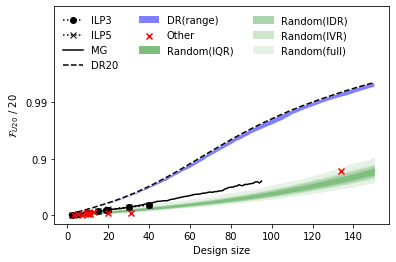}
\caption{For each vaccine design $S$ we compute $\objf{U_{T}}(S)$, divide it by $T$, and plot it as a function of $|S|$. Designs for both MHC Class I (A-D) and MHC Class II (E-H) are given. For each size we plot the entire range of designs of that size generated using the \dmr{} framework optimized for $\objf{U_{X}}$ with $1 \leq X \leq 20$ (DR(range)), and the interquartile (IQR), interdecile (IDR), interventile (IVR), and full range of 1000 designs of that size sampled uniformly at random. We also plot the locations of the designs generated by optimizing for $\objf{U_{T}}$ specifically (DR$T$), the locations of the ILP n=3 (ILP3), ILP n=5 (ILP5), and MarginalGreedy (MG) designs from \citetAppdx{ntimescoverage} as well as 29 other designs (Other) \citeAppdx{abdelmageed2020design,ahmed2020preliminary,akhand2020genome,alam2020design,banerjee2020energetics,baruah2020immunoinformatics,bhattacharya2020development,fast2020potential,gupta2020identification,herst2020effective,lee2020silico,mitramulti,poran2020sequence,ramaiah2021insights,saha2020silico,singh2020designing,srivastava2020structural,tahir2020designing,vashi2020understanding}.
Designs were evaluated using $\objf{U_1}$ (A,E), $\objf{U_3}$ (B,F), $\objf{U_8}$ (C,G), and $\objf{U_{20}}$ (D, H).}
\label{fig2_appdx_figevals}
\end{figure}

Figure \ref{fig2_comparedesigns} compares designs generated via the \dmr{} framework against previous designs using $\objf{U_5}$. Additional comparisons using $\objf{U_1}$, $\objf{U_3}$, $\objf{U_8}$, and $\objf{U_20}$ are provided in Figure \ref{fig2_appdx_figevals}. We similarly observe that the \dmr{} based designs achieve the best performance.

We also provide the $\objf{U_5}$ scores for the benchmark designs in Table \ref{table_main}, with comparisons against \dmr{} designs of matching size.

\begin{table}
\centering
\begin{tabular}{l|p{30mm}|p{15mm}|p{50mm}}

Peptide vaccine design (MHC1) & Vaccine size \newline (number of peptides) & $\objf{U_5}$ &
$\objf{U_5}$ evaluated on the \dmr{} design of the same size optimized for $\objf{U_5}$ \\
\hline
\citetAppdx{herst2020effective} & 52 & 3.76697 & 4.99580 \\
\citetAppdx{ntimescoverage} & 19 & 3.39579 & 4.69175 \\
\citetAppdx{ntimescoverage} & 19 & 3.20910 & 4.69175 \\
\citetAppdx{fast2020potential} & 13 & 2.93663 & 4.16200 \\
\citetAppdx{srivastava2020structural} & 37 & 2.14414 & 4.97735 \\
\citetAppdx{poran2020sequence} & 10 & 1.80712 & 3.63369 \\
\citetAppdx{herst2020effective} top16 & 16 & 1.68352 & 4.49048 \\
\citetAppdx{ahmed2020preliminary} & 16 & 1.46971 & 4.49048 \\
\citetAppdx{lee2020silico} & 13 & 1.34565 & 4.16200 \\
\citetAppdx{vashi2020understanding} & 51 & 1.12589 & 4.99536 \\
\citetAppdx{abdelmageed2020design} & 10 & 0.95321 & 3.63369 \\
\citetAppdx{akhand2020genome} & 31 & 0.91429 & 4.95097 \\
\citetAppdx{gupta2020identification} & 7 & 0.69957 & 2.82761 \\
\citetAppdx{baruah2020immunoinformatics} & 5 & 0.64483 & 2.12411 \\
\citetAppdx{mitramulti} & 9 & 0.57102 & 3.39366 \\
\citetAppdx{alam2020design} & 3 & 0.49469 & 1.33187 \\
\citetAppdx{saha2020silico} & 5 & 0.48918 & 2.12411 \\
\citetAppdx{bhattacharya2020development} & 13 & 0.41408 & 4.16200 \\
\citetAppdx{singh2020designing} & 7 & 0.40922 & 2.82761 \\

\hline\hline
Peptide vaccine design (MHC2) & Vaccine size \newline (number of peptides) & $\objf{U_5}$ &
$\objf{U_5}$ evaluated on the \dmr{} design of the same size optimized for $\objf{U_5}$ \\
\hline
\citetAppdx{ramaiah2021insights} & 134 & 4.99516 & 4.99988 \\
\citetAppdx{ntimescoverage} & 19 & 3.28835 & 4.69175 \\
\citetAppdx{ntimescoverage} & 19 & 3.25892 & 4.69175 \\
\citetAppdx{fast2020potential} & 13 & 2.54511 & 4.16200 \\
\citetAppdx{vashi2020understanding} & 20 & 1.95264 & 4.73921 \\
\citetAppdx{abdelmageed2020design} & 10 & 1.90814 & 3.63369 \\
\citetAppdx{akhand2020genome} & 31 & 1.72526 & 4.95097 \\
\citetAppdx{banerjee2020energetics} & 9 & 1.71810 & 3.39366 \\
\citetAppdx{poran2020sequence} & 10 & 1.10991 & 3.63369 \\
\citetAppdx{mitramulti} & 5 & 0.96545 & 2.12411 \\
\citetAppdx{tahir2020designing} & 11 & 0.77791 & 3.83742 \\
\citetAppdx{ahmed2020preliminary} & 5 & 0.55950 & 2.12411 \\
\citetAppdx{singh2020designing} & 7 & 0.40133 & 2.82761 \\
\citetAppdx{baruah2020immunoinformatics} & 3 & 0.00000 & 1.33187 \\

\end{tabular}
\caption{33 prior designs evaluated on $\objf{U_5}$, compared against designs generated by optimizing for $\objf{U_5}$ using the \dmr{} framework (Algorithm \ref{alg-greedy}).}
\label{table_main}
\end{table}

\subsection{Comparison of designs using n-times coverage}

\begin{figure}
\centering
A)\includegraphics[width=0.45\textwidth]{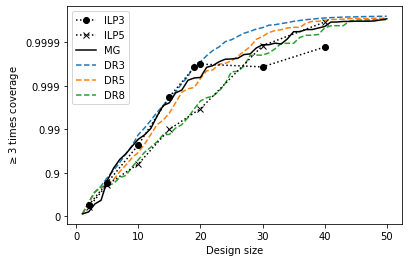}
D)\includegraphics[width=0.45\textwidth]{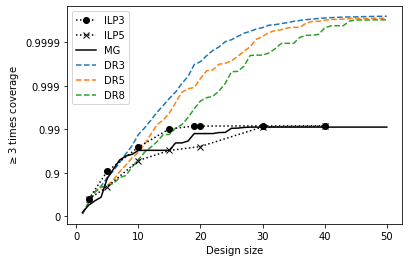}
B)\includegraphics[width=0.45\textwidth]{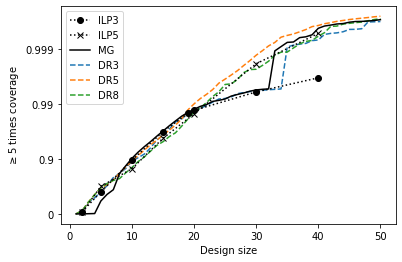}
E)\includegraphics[width=0.45\textwidth]{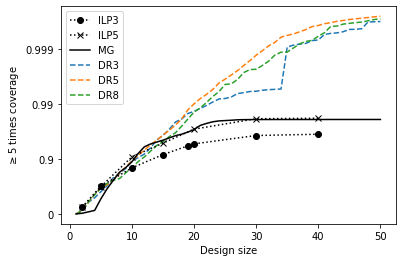}
C)\includegraphics[width=0.45\textwidth]{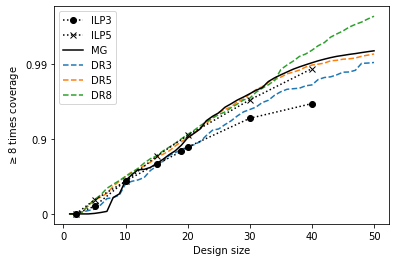}
F)\includegraphics[width=0.45\textwidth]{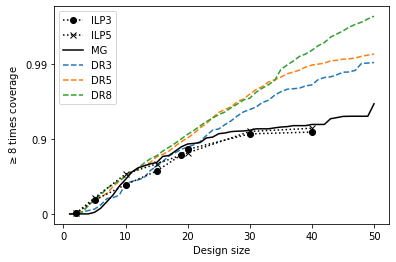}
\caption{For each vaccine design $S$ we compute the $n$-times coverage of $S$ and plot it as a function of $|S|$. Designs for both MHC Class I (A-C) and MHC Class II (D-F) are given. For each size we plot the $n$-times coverage of designs generated via the \dmr{} framework optimized via $\objf{U_3}$ (DR3), $\objf{U_5}$ (DR5), and $\objf{U_8}$ (DR8), as well as the $n$-times coverage of the ILP n=3 (ILP3), ILP n=5 (ILP5), and MarginalGreedy (MG) designs from \citetAppdx{ntimescoverage}. $n$-times coverage was computed for $n=3$ (A,D), $n=5$ (B,E), and $n=8$ (C,F).}
\label{fig2_appdx_figntimescover}
\end{figure}

We compare the performance of vaccines with 19 peptides designed from the \dmr{} framework against vaccines with 19 peptides from \citetAppdx{ntimescoverage} in Figure \ref{fig2_compareevalvax}. We additionally compare the performance of designs of varying sizes in Figure \ref{fig2_appdx_figntimescover}. The credences used in the objective for optimizing the \dmr{} designs were binarized to closely match the binding data used by \citetAppdx{ntimescoverage}, while the $n$-times coverage evaluation was reproduced from \citetAppdx{ntimescoverage}.

Similarly to our analysis from Figure \ref{fig2_compareevalvax}, we find that \dmr{} designs are highly competitive even when evaluated on the $n$-times coverage objective. We find that this is especially the case for designs for Class II MHCs, where prior approaches appear to get stuck at a certain thresholds.

\newpage

\bibliographystyleAppdx{plainnat}
\bibliographyAppdx{references}

\end{document}